\documentclass[letterpaper,twocolumn,superscriptaddress,prx]{revtex4-2}

\usepackage{graphicx}
\usepackage[colorlinks = true, citecolor=blue, linkcolor=black,urlcolor=purple]{hyperref}
\usepackage{mathtools,amssymb,amsmath}
\usepackage{cancel}
\usepackage{tikz}
\usetikzlibrary{quantikz2}

\newcommand{\tr}{\mathrm{tr}}
\newcommand{\HH}{{H}}
\newcommand{\UU}{{U}}
\newcommand{\RR}{{\rho}}
\newcommand{\cc}{{c}}
\newcommand{\cd}{{c}^{\dag}}
\newcommand{\gam}{{\gamma}}
\newcommand{\sig}{{\sigma}}
\newcommand{\LL}{{L}}
\newcommand{\PP}{{\mathcal{P}}}

\newcommand{\ibmyorktown}{IBM Quantum, IBM T.J. Watson Research Center, Yorktown Heights, 10598, USA}
\newcommand{\ibmcambridge}{MIT-IBM Watson AI Lab,  Cambridge MA, 02142, USA}
\newcommand{\UIUC}{Department of Physics, University of Illinois Urbana-Champaign, Urbana, Illinois 61801, USA}
\newcommand{\Hannover}{Institute of Quantum Optics, Leibniz Universität Hannover, Welfengarten 1, 30167 Hannover, Germany}

\newcommand{\ourtitle}{Quantum critical dynamics and emergent universality in decoherent digital quantum processors}

\begin{document}

\title{\ourtitle}
\author{Brendan Rhyno}
\email{rhyno@iqo.uni-hannover.de}
\affiliation{\Hannover}
\affiliation{\UIUC}
\author{Swarnadeep Majumder}
\affiliation{\ibmyorktown}
\affiliation{\ibmcambridge}
\author{Smitha Vishveshwara}
\affiliation{\UIUC}
\author{Khadijeh Najafi}
\email{sonaa.najafi@gmail.com}
\affiliation{\ibmcambridge}
\affiliation{\ibmyorktown}

\maketitle

\textbf{Understanding how noise influences nonequilibrium quantum critical dynamics is essential for both fundamental physics and the development of practical quantum technologies.
While the quantum Kibble-Zurek (QKZ) mechanism predicts universal scaling during quenches across a critical point, real quantum systems exhibit complex decoherence that can substantially modify these behaviors, ranging from altering critical scaling to completely suppressing it.
By considering a specific case of nondemolishing noise, we first show how decoherence can reshape universal scaling and verify these theoretical predictions using numerical simulations of spin chains across a wide range of noise strengths.
Then, we study linear quenches in the transverse-field Ising model on IBM superconducting processors where the noise model is unknown.
Using large system sizes of 80-120 qubits, we measure equal-time connected correlations, defect densities, and excess energies across various quench times.
Surprisingly, unlike earlier observations where noise-induced defect production masked universal behavior at long times, we observe clear scaling relations, pointing towards persistent universal structure shaped by decoherence. The extracted scaling exponents differ from both ideal QKZ predictions and analytic results for simplified noise models, suggesting the emergence of a distinct noise-influenced universality regime. Our results, therefore, point toward the possibility of using universal dynamical scaling as a high-level descriptor of quantum hardware, complementary to conventional gate-level performance metrics.}

\vspace{0.5cm}
{\textbf{Introduction}}

\begin{figure*}[t!]
    \centering
    \includegraphics[width=0.9\textwidth]{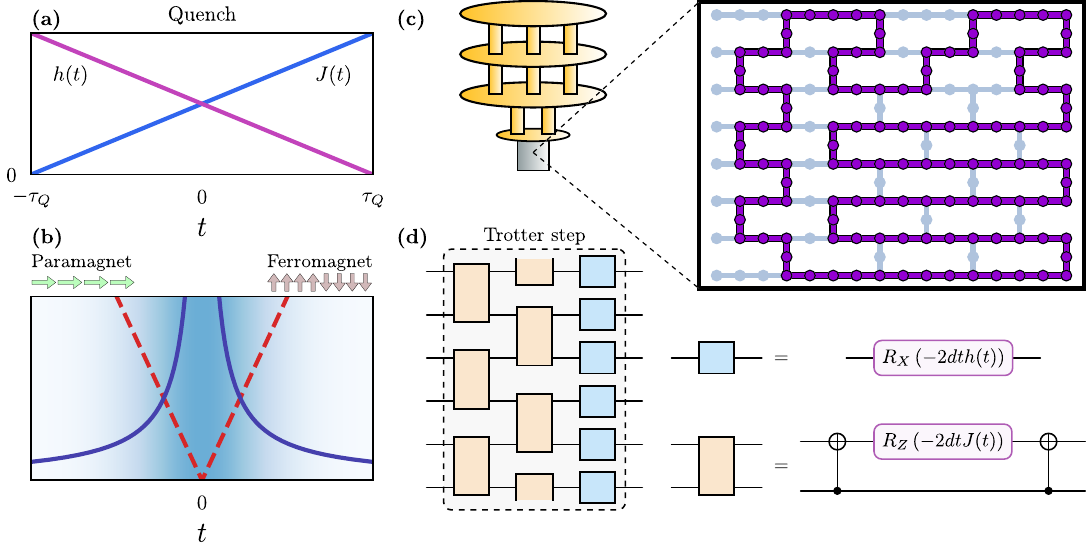}
    \caption{
    \textbf{Digital quantum simulations of the quantum Kibble-Zurek mechanism.}
    (a)
    Schematic of a continuous linear quench of the one-dimensional transverse-field Ising model (TFIM).
    The Hamiltonian is ramped according to $J(t) = ( 1 + t / \tau_Q )$ and $h(t) = ( 1 - t/\tau_Q )$, with $\tau_Q$ denoting the characteristic quench time.
    The evolution starts from the paramagnetic ground state at $t = -\tau_Q$, crosses the quantum critical point at $t=0$, and reaches the Ising Hamiltonian at $t=\tau_Q$.
    (b)
    Near the critical point, the control parameter becomes vanishingly small, $|\epsilon(t)| \propto |t / \tau_Q |$ as $t \to 0$ (dashed red line), and the correlation time, $\xi_t \propto |\epsilon|^{-z \nu}$, diverges (solid blue line).
    The Kibble-Zurek freeze-out regime associated with the quench ramp is indicated by the light-blue shaded region.
    In this regime, the correlation length, $\xi \propto |\epsilon|^{-\nu}$, grows rapidly, but as a result of the critical slowing down finite-sized domains become effectively frozen and the system loses its ability to suppress defect formation.
    (c)
    An example qubit layout on the \texttt{ibm\_fez} Heron processor for a chain with 120 sites.
    (d)
    Trotterization of the quench protocol for the TFIM.
    The unitary time-evolution operator is implemented using a first-order Suzuki-Trotter decomposition with discrete time step $ dt $.
    The on-site transverse-field term maps to a layer of single-qubit $ R_X $ rotation gates, while the nearest-neighbor Ising interaction is constructed from two-qubit unitaries involving two CNOT gates and a $ R_Z $ rotation gate.
    To reduce qubit idling and enable parallel execution, the interaction gates are applied in sequential ``odd'' and ``even'' sublayers.
    }
    \label{fig:Experimental_setup}
\end{figure*}

Understanding thermalization and nonequilibrium dynamics in quantum systems remains as one of the fundamental problems in many-body quantum systems.
The Kibble-Zurek (KZ) mechanism stands as a cornerstone in the study of phase transitions, offering a universal description for the formation of topological defects when a system is driven through a critical point at a finite rate~\cite{Kibble1976,Kibble1980,Zurek1985,Zurek1996}.
As a zero-temperature system is driven across a quantum critical point and the many-body gap closes, it cannot adjust adiabatically and its state ``freezes,'' leading to the formation of defects.
The quantum Kibble-Zurek (QKZ) mechanism quantifies this process by linking the defect density to the quench rate and the systems critical exponents, see Fig.~\ref{fig:Experimental_setup}(b),~\cite{Zurek2005,Dziarmaga2005,Polkovnikov2005,Schuetzhold2006,Uhlmann2007,Uhlmann2010,Dziarmaga2010,Polkovnikov2011}.

This framework has gained particular importance with the rise of quantum technologies such as quantum computing and simulation~\cite{Cirac2012,Daley2022,Fauseweh2024,Bloch2008,Blatt2012,Houck2012,Lewenstein2012,Smith2016,Zhang2017,Bernien2017,Smith2019,Browaeys2020,Bluvstein2021,Monroe2021,Mi2022,Vezvaee2024,Farrell2024,King2025,Minev2025,Gyawali2025,Hayata2025,Koyluouglu2026}, where accurate control of quantum states during phase transitions is essential for advancing device capabilities.
Recent experimental demonstrations of the QKZ mechanism across diverse platforms, including Rydberg arrays, superconducting qubits, and ultra-cold atoms, underscore its broad applicability and significance~\cite{Keesling2019,Ebadi2021,Dupont2022,Miessen2024,Manovitz2025,Andersen2025}.
Yet, a critical challenge emerges in real-world systems: the pervasive presence of noise, arising from environmental interactions, imperfect control, or measurement imperfections, which can perturb the ideal dynamics predicted by the QKZ mechanism and alter the resulting universal scaling~\cite{Miessen2024}.

Understanding how noise affects the QKZ mechanism is essential not only for validating its predictions under realistic conditions but also for advancing practical quantum algorithms, including quantum annealing and adiabatic state preparation~\cite{Kato1950,Avron1999,Suzuki2007}.
It has been shown that decoherence, measurement, and environmental coupling introduce additional timescales that fundamentally reshape nonequilibrium dynamics.
These results underscore that the interplay between coherent dynamics and dissipation or noise can preserve, distort, or invert KZ scaling, depending on the intrinsic nature of the noise and its strength, highlighting both the fragility and adaptability of universal dynamics in noisy quantum systems
~\cite{Miessen2024,Keck2017,Kuo2021,Griffin2012,Dutta2016,Gao2017,Puebla2020,Gangadharaiah2021,Singh2023,Iwamura2024}.
In particular, in the case of nondemolishing noise, the emergence of universality has been predicted with modified critical exponents~\cite{Kuo2021}.

Although significant progress has been made toward understanding the impact of noise on critical dynamics, a complete characterization remains elusive, especially in experimentally accessible quantum platforms. To address this gap, we investigate the role of noise in the QKZ mechanism using digital superconducting quantum processors. Our results are consistent with emergent universal behavior, manifested through the collapse of connected two-point correlation functions across a wide range of quench times and system sizes, along with scaling relations for the defect density and excess energy produced during the quench.
Notably, we find that the associated exponents are modified compared to both the standard QKZ predictions and the values reported under nondemolishing noise in Ref.~\cite{Kuo2021}.
Our observations extend beyond theoretical predictions and noise models that are tractable with classical simulations.

\vspace{0.5cm}
{\textbf{Experimental setup}}

To this end, we run our simulations on a 156-qubit \texttt{ibm\_fez} Heron processor which uses fixed-frequency transmon qubits as illustrated in Fig.~\ref{fig:Experimental_setup}(c).
Here, we start with the well-known one-dimensional transverse-field Ising model (TFIM) defined as 
\begin{equation}
    \label{eq:TFIM}
    H(t)
    =
    - J(t) \sum_{i=1}^{N} \sig^z_i \sig^z_{i+1}
    - h(t) \sum_{i=1}^{N} \sig^x_i
    ,
\end{equation}
with $J(t) \ge 0$ and $h(t)\ge 0$ and $\sig^\alpha_{N+1} \equiv \sig^\alpha_{1}$ employing periodic boundary conditions. The dimensionless parameter $\epsilon=1-h/J$ controls the zero temperature physics of the model and when $\epsilon = 0$ there is a quantum phase transition in the thermodynamic limit ($N\to\infty$)~\cite{Sachdev2001}.

The system is initially prepared in the paramagnetic ground state $|+\rangle^{\otimes N}$.
We then subject the system to a linear quench protocol that is suitable for both numerical calculations and quantum-hardware implementations. In particular, we adopt the scheme used in Refs.~\cite{Dupont2022, Miessen2024}, where the time-dependent couplings evolve as $J(t) = ( 1 + t / \tau_Q )$ and $h(t) = ( 1 - t / \tau_Q )$, with $\tau_Q$ denoting the characteristic quench time.
Throughout, we adopt units where $\hbar = J(0) = 1$.
In this work, we explore two variants of the protocol: (A) a quench carried out over the interval $-\tau_Q \leq t \leq 0$ evolving the system to its quantum critical point (QCP); and (B) an extended quench that drives the system fully into the ferromagnetic phase by spanning the entire interval $-\tau_Q \leq t \leq \tau_Q$. 
The corresponding setup is illustrated in Fig.~\ref{fig:Experimental_setup}(a).

Near a QCP, universal scaling laws govern the dynamics. The correlation length diverges as $\xi \propto |\epsilon|^{-\nu}$ and the energy gap closes as $\Delta \propto |\epsilon|^{z\nu}$, with $\nu$ and $z$ the correlation length and dynamical exponents, respectively ($z=\nu=1$ for the one-dimensional TFIM). During a linear quench, the control parameter varies as $\epsilon(t) \propto t/\tau_Q$ as $t \to 0$.
Although adiabatic evolution is maintained far from criticality, it necessarily fails near the QCP due to critical slowing down (see Fig.~\ref{fig:Experimental_setup}(b)). This breakdown defines a freeze-out time $\bar{t} \propto \tau_Q^{z\nu/(1+z\nu)}$, at which the correlation length saturates to $\bar{\xi} \propto \tau_Q^{\nu/(1+z\nu)}$, setting the characteristic size of equilibrium domains.
The Kibble-Zurek scaling hypothesis predicts that, in the limit $ \tau_Q \to \infty $, $ \bar{\xi} $ and $ \bar{t} $ become the only relevant length and time scales in the problem~\cite{Francuz2016}.

To investigate the dynamics on quantum processors, we discretize time into segments of equal duration $dt$ and step the system forward in time with the unitary time-evolution operator $U(t) = \exp(-i dt H(t))$.
We implement this operator by employing a first-order Suzuki-Trotter decomposition~\cite{Trotter1959,Suzuki1993,Lloyd1996,Zalka1998} (see Supplementary Material~\ref{SI:quantum_simulation}).
Although the Ising and transverse-field terms in Eq.~(\ref{eq:TFIM}) do not commute, the local terms within each operator commute among themselves. The on-site transverse-field term maps directly onto a layer of single-qubit $ R_X $ rotation gates, whereas the nearest-neighbor interaction term is implemented using a standard construction with a $ R_Z $ rotation gate sandwiched between two CNOT gates~\cite{Smith2019}.
To minimize qubit idling during gate execution, the interaction circuits are applied in ``odd'' and ``even'' sublayers, as illustrated in Fig.~\ref{fig:Experimental_setup}(d).
This staggered application enables efficient parallelization of entangling operations across the quantum register.
After preparing the initial state and time evolving the system to a desired discrete time, $t_s = -\tau_Q + s dt$ with $s$ an integer, the qubits are measured along an appropriate axis (either $x$ or $z$). Upon many such realizations, equal-time spin correlation functions can be constructed statistically from the shot data.
All experiments reported in this work utilized $2^{20}$ shots, which leads to standard errors in the correlation functions on the order of $10^{-3}$ (see Supplementary Material~\ref{SI:sampling}).

\vspace{0.5cm}
{\textbf{Local magnetization and connected correlations during the quench}}

\begin{figure*}[t]
    \centering
    \includegraphics[width=0.9\textwidth]{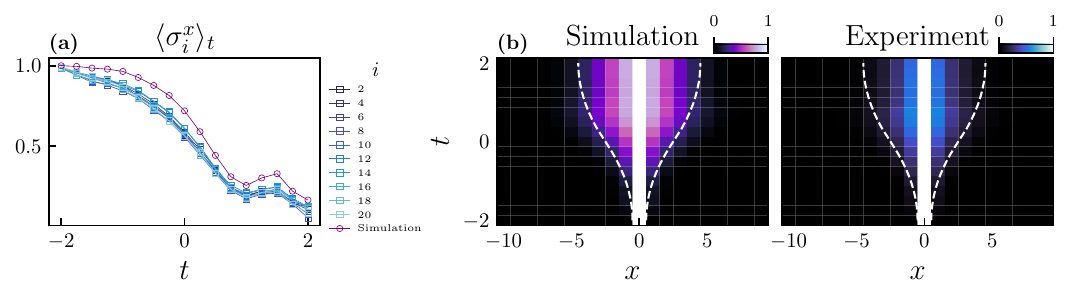}
    \caption{
    \textbf{Local magnetization and connected correlations during a quench of the transverse-field Ising model.}
    Experimental measurements and theoretical simulations are shown for a quench from the paramagnetic ground state into the ferromagnetic phase, implemented using $ N=20 $ qubits and 16 Trotter steps of duration $ dt=0.25 $ ($ -2 \le t \le 2 $).
    (a)
    Measurements along the $ x $ axis yield the local magnetization $ \langle \sig_i^x \rangle_t $ at each discrete time step (blue squares).
    The experimental data closely follow the theoretical evolution of a closed system undergoing the Trotterized quench (purple circles), with residual deviations attributable to hardware noise.
    (b)
    Measurements along the $ z $ axis yield the equal-time connected spin--spin correlation function $ |C(t,x)| $.
    Experimental results are shown in the right panel and zero-noise theoretical simulations in the left panel.
    The dashed white lines indicate the light-cone trajectories for the spread of correlations, computed in the thermodynamic limit for the continuous quench protocol shown in Fig.~\ref{fig:Experimental_setup}(a).
    }
    \label{fig:N=20_Dynamics}
\end{figure*}

First, using a chain of $ N=20 $ qubits implemented on the \texttt{ibm\_fez} processor (see Fig.~\ref{fig:Experimental_setup}(c)), we demonstrate time evolution from the initial paramagnetic Hamiltonian to the final ferromagnetic Hamiltonian by measuring the system along the $x$-axis and extracting the $x$-component of the local magnetization, $ \langle \sig^x_i \rangle_t$. 

The experiments were carried out with a quench time of $ \tau_Q=2 $ achieved using 16 Trotter steps of duration $ dt=0.25 $ over the interval $-\tau_Q \le t \le \tau_Q$.
In Fig.~\ref{fig:N=20_Dynamics}(a), we see that all the spins point up along the $x$-axis in the initial paramagnetic phase. However, as the quench progresses into the ferromagnetic phase oriented along the $z$-axis, this initial magnetization, depicted as blue squares for different sites, gets strongly suppressed. In addition, as expected, we notice a deviation from the noiseless simulation depicted with purple circles. 
To further investigate the effects of hardware noise, in Fig.~\ref{fig:N=20_Dynamics}(b), we demonstrate the spread of correlations in the system using the (site-averaged) equal-time connected spin-spin correlation function:
\begin{equation}
    \label{eq:site_averaged_connected_2pt_fn}
    C(t,x)
    =
    \frac{1}{N} \sum_{i=1}^{N}
    \left[
    \langle \sig^z_i \sig^z_{i+x} \rangle_t - \langle \sig^z_i \rangle_t \langle \sig^z_{i+x} \rangle_t
    \right]
    ,
\end{equation}
where $x$ denotes the spatial separation.
As the system is quenched toward the critical point, correlations begin to spread from the central site and increase from zero, reflecting the transition from the paramagnetic state. For comparison, we also include results from a noiseless simulation using the Trotterized time evolution operator in the middle panel of the figure. In small systems, we observe excellent agreement between the experimental data and exact noiseless simulations. However, the spread of correlations becomes noticeably suppressed in the experimental data at longer times due to noise in the hardware, as depicted in the rightmost panel.

\vspace{0.5cm}
{\textbf{Universal scaling under quantum nondemolishing noise}}

\begin{figure*}[t]
    \centering
    \includegraphics[width=0.9\textwidth]{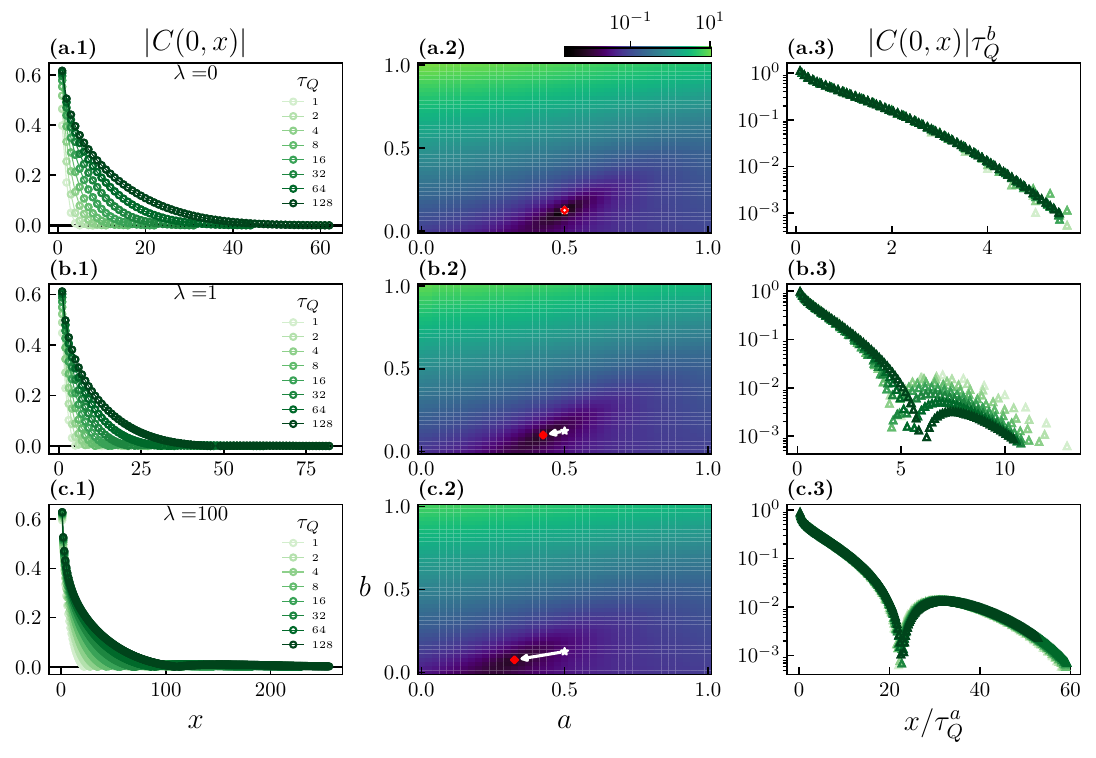}
    \caption{
    \textbf{Scaling behavior at the quantum critical point under quantum nondemolition noise.}
    Numerical simulations of $N=512$ spins subject to the continuous quench protocol to the QCP, shown in Fig.~\ref{fig:Experimental_setup}(a), for various quench times and QND noise strength values.
    Rows (a)-(c) show results after solving Eq.~(\ref{eq:Master_eq}) for quenches performed with $\lambda=0,1,100$ respectively.
    The correlation function, Eq.~(\ref{eq:site_averaged_connected_2pt_fn}), is depicted in column 1 (values below $5\times 10^{-4}$, as well as the $x=0$ point, are masked throughout).
    In column 2 we show RMSE results for different candidate scaling exponents, $(a,b)$, using the search protocol described in the main text.
    The scaling exponents predicted by the standard (noiseless) QKZ mechanism are marked with white stars and those found by the search algorithm under decoherence strength $\lambda$ are marked with red diamonds.
    Using the exponent pair with the lowest RMSE, in column 3 we present the rescaled correlation function data, $C(0,x) \tau_Q^{b_\mathrm{fit}}$ vs $x / \tau_Q^{a_\mathrm{fit}}$.
    (a) and (c) In the two extreme limits of zero- and large-QND noise strength, the exponent pair with the lowest RMSE corresponds (within the grid precision of $0.025$) to the theoretical predictions of $(a_\mathrm{QKZ},b_\mathrm{QKZ})=(1/2, 1/8)$ and $(a_\mathrm{QND},b_\mathrm{QND})=(1/3,1/12)$ respectively.
    In both cases, there is a noticeable data collapse in the rescaled correlation functions.
    (b) At intermediate QND noise strengths, where scaling relations are not predicted to hold, the fitted exponents do not yield a clear data collapse.
    }
    \label{fig:QND_QKZ}
\end{figure*}

The QKZ mechanism predicts that the equal-time connected spin-spin correlation function satisfies the scaling form $C(t,x) = \bar\xi^{-\Delta_{zz}} \mathcal{F} (t/\bar t , |x| / \bar\xi ) $
, where $\Delta_{zz}$ is the scaling dimension of the operator $\sig^z_i \sig^z_j$ and $\mathcal{F}$ is a non-universal function of the rescaled time and distance.
Expressing the scaling hypothesis in terms of the quench time $\tau_Q$, the QKZ hypothesis predicts a \textit{data collapse} for correlation functions at the QCP ($t=0$) across a range of quench times when both $C(0,x)$ and $|x|$ are rescaled according to
\begin{equation}
    \label{eq:C_QKZ}
    C(0,x) \tau_Q^{b_\mathrm{QKZ}} = \mathcal{G} ( |x| / \tau_Q^{a_\mathrm{QKZ}} )
    ,
\end{equation}
where $a_\mathrm{QKZ} \equiv \nu / (1 +  z \nu)$ and $b_\mathrm{QKZ} \equiv \Delta_{zz} \nu / (1 + z \nu)$.
For the one-dimensional Ising universality class, $z = \nu = 1$ and $\Delta_{zz} = 1/4$, yielding $a_{\mathrm{QKZ}} = 1/2$ and $b_{\mathrm{QKZ}} = 1/8$.
Data collapse in this model has been demonstrated in the thermodynamic limit for sufficiently slow quenches~\cite{Francuz2016}.

A growing body of work has examined how nonequilibrium scaling behaves in the presence of noise, measurements, or environmental coupling. 
In such scenarios, new dynamical scales emerge which can compete with or even supersede the intrinsic KZ timescales.
Depending on the nature and strength of the noise, the resulting dynamics may continue to display KZ-like behavior, follow modified scaling relations, or even exhibit inverted trends~\cite{Miessen2024,Keck2017,Kuo2021,Griffin2012,Dutta2016,Gao2017,Puebla2020,Gangadharaiah2021,Singh2023,Iwamura2024}.
These findings demonstrate that universal dynamics in quantum systems are highly sensitive to the interplay between coherent evolution and dissipative processes.
In particular, when the noise originates from stochastic control fields, as explored in~\cite{Dutta2016}, the system may exhibit nonmonotonic dependence on the quench time where slower quenches result in more excitations.
This \textit{anti}-KZ behavior is a complete inversion of the standard QKZ expectations~\cite{Griffin2012,Dutta2016,Gao2017,Puebla2020,Gangadharaiah2021,Singh2023,Iwamura2024}.
In digital simulations, these effects become intertwined with gate noise and circuit depth limitations. Yet, as shown in recent benchmarking work, KZ dynamics still manifest robustly in quantum simulators with hundreds of qubits, albeit with quantitative deviations due to device imperfections~\cite{Miessen2024}.

As a concrete example of noise-induced modifications to QKZ scaling, we consider the effect of continuous quantum nondemolition (QND) measurements of the instantaneous Hamiltonian. In this setting, the system undergoes decoherence in the instantaneous energy eigenstates, leading to modified freeze-out behavior in the strong decoherence regime and new algebraic combinations of universal critical exponents in the freeze-out time and correlation length~\cite{Kuo2021}:
\begin{equation}
    \label{eq:QND_freezeout}
    \bar{t} \propto \tau_Q^{2 z \nu /(1 + 2 z \nu)}
    \quad,\quad
    \bar{\xi} \propto \tau_Q^{\nu /(1 + 2 z \nu)}
    .
\end{equation}
Similar to Eq.~(\ref{eq:C_QKZ}), a data collapse at the QCP is predicted, but now with exponents $a_\mathrm{QND} \equiv \nu / (1 + 2 z \nu)=1/3$ and $b_\mathrm{QND} \equiv \Delta_{zz} \nu / (1 + 2 z \nu)=1/12$ for the one-dimensional Ising universality class.

To examine the effect of continuous measurements, in Fig.~\ref{fig:QND_QKZ} we present our results under varying the strength of nondemolishing quantum noise in the TFIM.
After mapping the TFIM onto spinless Jordan-Wigner fermions, the dynamics of the system under nondemolishing quantum noise can be approximated by solving a master equation
(we have provided the full derivation in the Supplementary Material~\ref{SI:QND})
\begin{equation}
    \label{eq:Master_eq}
    \partial_t \vec{n}_k(t)
    =
    \vec{B}_k(t) \times \vec{n}_k(t)
    +
    \lambda \, \vec{B}_k(t) \times (\vec{B}_k(t) \times \vec{n}_k(t))
    ,
\end{equation}
where $\vec{n}_k(t)$ is a Bloch vector for each momentum mode $k$ with $|\vec{n}_k(t)| \le 1$ and $\vec{B}_k(t)$ is a pseudo-magnetic field in Nambu space.
Here, $\lambda \ge 0$ controls the strength of nondemolishing measurements.
After solving the master equation numerically for each $k$ mode, we extract 2-point fermionic correlation functions, from which we construct the equal-time connected spin-spin correlation function $C(t,x)$, shown in the first column of Fig.~\ref{fig:QND_QKZ}.

For both the numerically generated correlation function data presented here and the experimental data to be discussed next, we use an optimization scheme similar to the one developed in~\cite{Dupont2022} to search for exponents $(a,b)$ that lead to the best data collapse.
As the exponents $a$ and $b$ are swept over a square-grid, the rescaled $C(0,x) \tau_Q^{b}$ versus $x / \tau_Q^{a}$ data are fit to an exponentially decaying function (see Supplementary Material~\ref{SI:fitting}).
The best-fit exponents, $(a_\mathrm{fit},b_\mathrm{fit})$, are taken to be the pair whose fit yields the smallest root mean square error (RMSE).

Sweeping the value of $\lambda$, in the second column of Fig.~\ref{fig:QND_QKZ}, we present the RMSE obtained after performing the fit for each pair of exponents $(a,b)$ on a coarse grid (our interest is in the flow of the exponents, not their precise values).
For $\lambda = 0$, we recover the standard zero-noise QKZ scaling, namely $a_\mathrm{fit}=0.5$ and $b_\mathrm{fit}=0.125$ with an excellent collapse visible in the rescaled data shown in the third column of Fig.~\ref{fig:QND_QKZ}(a).
At intermediate QND noise strengths such as $\lambda=1$, universal scaling is not expected to hold, consequently we observe that the best-fit exponents do not produce a convincing collapse, see Fig.~\ref{fig:QND_QKZ}(b).
For a large QND noise strength of $\lambda=100$, the scaling exponents converge (within the precision of the grid) to those predicted in the $\lambda \to \infty$ limit, $(a_\mathrm{QND},b_\mathrm{QND})=(1/3,1/12)$, and again yield a pronounced data collapse.
Our numerics show a clear crossover in the universal scaling behavior from noiseless QKZ predictions (white stars in the figure) to the strong decoherence regime predicted in~\cite{Kuo2021}.

\vspace{0.5cm}
{\textbf{Scaling behavior at the critical point under the decoherence of quantum processors}}

\begin{figure*}[t]
    \centering
    \includegraphics[width=0.9\textwidth]{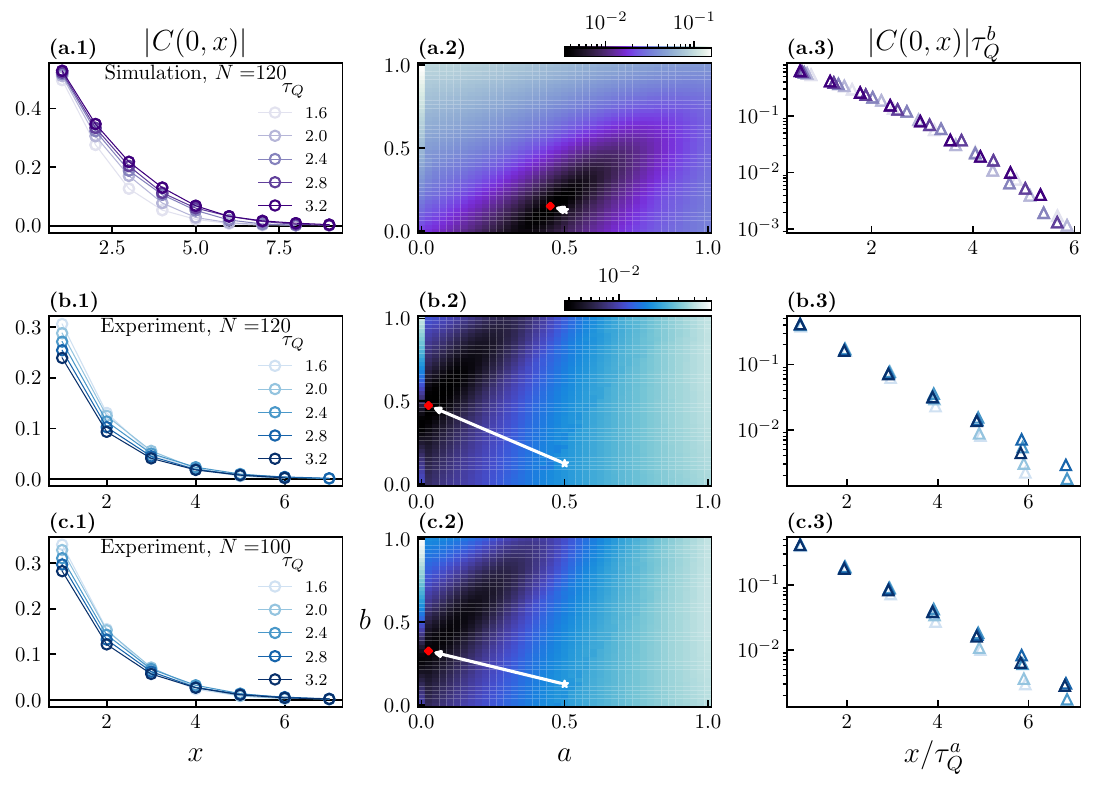}
    \caption{
    \textbf{Scaling behavior at the quantum critical point in digital quantum computers.}
    (a)
    Numerical simulations of $N=120$ spins undergoing a Trotterized quench to the QCP for various quench times ($\tau_Q = d t \cdot \mathrm{steps}$ with $d t = 0.2$). 
    (b) and (c)
    Digital quantum simulations of the Trotterized quenches in (a) performed on \texttt{ibm\_fez} using 120 and 100 qubits respectively.
    Similar to Fig.~\ref{fig:QND_QKZ}, column 1 shows the correlation function data as function of distance, column 2 the search for scaling exponents (on a grid with spacing $0.025$), and column 3 the rescaled correlation function data using the best-fit exponent pair.
    Because of statistical shot noise, correlation function data below $10^{-3}$ is omitted, as is the $x=0$ point where QKZ scaling does not apply.
    (a)
    In the theory simulation, Trotterization and finite-size effects lead to small deviations between the fitted scaling exponents $(a_\mathrm{fit},b_\mathrm{fit})=(0.45, 0.15)$ (red diamond) and those predicted by the QKZ mechanism (white star).
    (b) and (c)
    Using IBM Quantum hardware, we observe significantly stronger deviations from QKZ scaling predictions. 
    Rather than correlations growing with longer quench times, they are suppressed.
    As expected, the extracted best-fit scaling exponents differ markedly from the QKZ hypothesis yielding 
    $(0.025,\,0.475)$ for $N=120$ and $(0.025,\,0.325)$ for $N=100$. However, despite the deviation, the fitted curves still exhibit a data collapse, consistent with the emergence of universal scaling behavior shaped by quantum decoherence.
    }
    \label{fig:Anti-KZ_QCP}
\end{figure*}

Next, we use our set-up to explore the signatures of universal scaling in the correlation functions at the critical point by simulating the quench protocol using the IBM Quantum processors.
In Fig.~\ref{fig:Anti-KZ_QCP}(b-c) we present experimental results obtained from quenching the system to the QCP using $N=100$ and $N=120$ qubits on \texttt{ibm\_fez} with Trotter steps of duration $ dt=0.2 $.
In order to distinguish the effect of the noise from Trotter error and finite-size effects, in Fig.~\ref{fig:Anti-KZ_QCP}(a), we first consider how Trotterized time evolution influences scaling in an ideal noiseless chain.
By numerically solving Eq.~(\ref{eq:Master_eq}) for a closed quantum system ($\lambda=0$) containing $N=120$ qubits, we obtain the connected spin-spin correlation function and search for the best scaling exponents to achieve data collapse in the $C(0,x) \tau_Q^{b}$ versus $x / \tau_Q^{a}$ data.
Although one obtains a convincing data collapse, the best-fit scaling exponents $(a_\mathrm{fit},b_\mathrm{fit})=(0.45,0.15)$ are already shifted from QKZ predictions in a noiseless system, $(a_\mathrm{QKZ},b_\mathrm{QKZ})=(0.5,0.125)$, due to the combination of Trotterization and finite-size effects.

In the case of quantum simulations performed on \texttt{ibm\_fez}, in Fig.~\ref{fig:Anti-KZ_QCP}(b-c) we observe a qualitatively different feature in the correlation function data as the quench time increases. In contrast to the zero-noise system, where the correlation function at a given distance tends to grow with quench time, in our quantum simulations, we observe a tendency for correlations to decrease with $\tau_Q$.
This behavior~\cite{Dutta2016} reflects the significant influence that the noisy environment has on the qubit dynamics.  This is further quantified by the strong shift in the best-fit scaling exponents using the \texttt{ibm\_fez} correlation function data compared to the standard QKZ result. For the $N=100$ and $N=120$ hardware runs we obtain $(0.025,0.325)$ and $(0.025,0.475)$, respectively. While this deviation superficially resembles the behavior predicted in the QND case, an important distinction must be made: unlike QND, we lack a theoretical prediction for real quantum processors. The noise landscape on current hardware is far more complex, arising from a convoluted combination of quantum decoherence, crosstalk, drift, and even heating effects. However, once the fitting procedure converges, we observe data collapse across different system sizes and experiments performed at different times, suggesting the possible emergence of universal behavior sustained by underlying quantum coherence.

\vspace{0.5cm}
{\textbf{Emergence of the anti-Kibble-Zurek mechanism in quantum processors}}

\begin{figure*}[t]
    \centering
    \includegraphics[width=0.6\textwidth]{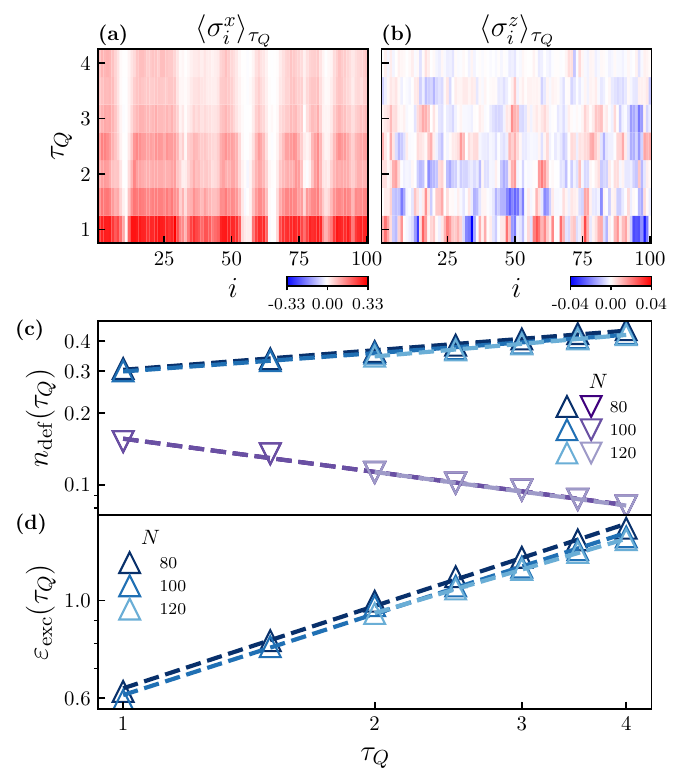}
    \caption{
    \textbf{Magnetization, defect density, and excess energy following a full quench of the transverse-field Ising model.}
    Digital quantum simulations were performed using $N=80, 100,$ and $120$ qubits on \texttt{ibm\_fez}.
    The system is measured at the end of the quench, $t=\tau_Q$, while varying the total number of Trotter steps between 8 and 32 at fixed step size $dt=0.25$.
    (a) and (b) Local magnetization profiles measured along the $x$ and $z$ axes, respectively, for different quench times in the $N=100$-qubit chain.
    The experimental data reveal deviations from translational invariance induced by hardware noise, as well as spatial structure in the longitudinal magnetization consistent with the formation of finite-sized domains.
    (c) and (d) Defect density $n_{\mathrm{def}}$ and excess energy $\varepsilon_{\mathrm{exc}}$ as functions of the quench time $\tau_Q$ for different system sizes (blue upward-pointing triangles).
    Power-law fits yield $n_{\mathrm{def}}(\tau_Q)\propto \tau_Q^{-\beta}$ with $\beta \approx - 0.3$ and $\varepsilon_{\mathrm{exc}}(\tau_Q)\propto \tau_Q^{-\gamma}$ with $\gamma \approx - 0.6$.
    These results are contrasted with theoretical simulations of the same Trotterized quench in a noiseless system (purple downward-pointing triangles), for which the defect density closely follows $n_{\mathrm{def}}(\tau_Q)\propto \tau_Q^{-1/2}$, in agreement with the exact result for a linear quench in the thermodynamic limit.
    }
    \label{fig:Anti-KZ_ISING}
\end{figure*}

To further elucidate the impact of noise on universal scaling, we next perform additional experiments in which the system is quenched to the ferromagnetic Hamiltonian, requiring analysis over the full time interval.
In Fig.~\ref{fig:Anti-KZ_ISING}(a-b) we show the local magnetization measured along the $x$- and $z$-axes for a $N=100$ qubit chain.
Consistent with the QKZ mechanism, paramagnetic order along the $x$-axis is generally suppressed for slower quenches into the ferromagnetic regime.
However, in contrast to a noiseless system, the observables $\langle \sig^x_i\rangle_{\tau_Q}$ and $\langle \sig^z_i\rangle_{\tau_Q}$ are not translation invariant in hardware simulations.
For instance, small fluctuations and domain-walls appear along the $z$-axis instead of the zero-noise result that $\langle \sig^z_i\rangle_{\tau_Q} = 0$.
In order to study signatures of universal behavior, we also measure the density of defects produced at the end of the quench by averaging over all edges of the spin lattice:
\begin{equation}
    \label{eq:defect_density}
    n_{\mathrm{def}}(\tau_Q)=\frac{1}{2N}\sum_{i=1}^{N}(1-\langle \sig^z_i \sig^z_{i+1} \rangle_{\tau_Q})
    .
\end{equation}
The QKZ mechanism predicts the scaling relation $ n_{\mathrm{def}}(\tau_Q) \propto \tau_Q^{-\beta} $ with $\beta \equiv d\nu/(1 + z\nu) = 1/2$ for the one-dimensional Ising universality class~\cite{Zurek2005,Dziarmaga2005,Polkovnikov2005,Dziarmaga2010,Polkovnikov2011,Dutta2016}.
In Fig.~\ref{fig:Anti-KZ_ISING}(c), we show the corresponding defect scaling on IBM Quantum hardware for $N=80$, $100$, and $120$ qubits.
While in~\cite{Miessen2024}, the authors reported deviations from KZ scaling due to finite-size effects and noise in the hardware, here, we detect a phenomenon known as the anti-KZ mechanism, a regime where slower quenches paradoxically result in more excitations~\cite{Griffin2012,Dutta2016,Gao2017,Puebla2020,Gangadharaiah2021,Singh2023,Iwamura2024}.
In particular, for all three system sizes we find $n_\mathrm{def}(\tau_Q) \propto \tau_Q^{-\beta}$ with $\beta\approx -0.3$, nearly inverting standard QKZ scaling.
Thus, for longer quenches we observe that the increased gate depth and noise experienced by the system leads to a growing number of defects.

In quenching to the ferromagnetic Hamiltonian, we also explore the excess energy density $ \varepsilon_{\text{exc}} $  which we define as the difference between the measured energy density of the system and the energy density of a closed (noiseless) quantum system~\cite{Keck2017}:
\begin{equation}
    \varepsilon_{\text{exc}}(\tau_Q) = \frac{1}{N} \Big( \tr\left[ \rho(\tau_Q) H(\tau_Q) \right] - \langle \psi(\tau_Q) | H(\tau_Q) | \psi(\tau_Q)\rangle \Big)
    ,
\end{equation}
where  $ \rho(\tau_Q) $  is the density matrix of the system at the final time, and $ | \psi(\tau_Q)\rangle = \mathcal{T}\exp\left[-i \int_{-\tau_Q}^{+\tau_Q}dt H(t) \right] | + \rangle^{\otimes N} $ is the final state of the system upon reaching the ferromagnetic Hamiltonian assuming unitary time evolution.
In contrast to the residual energy, which subtracts off the instantaneous ground state energy and quantifies the system's failure to remain in its ground state due to non-adiabatic transitions, the excess energy further quantifies deviations in the measured energy due to the system's interactions with its environment.
[Note: for the quench protocol considered in this work, the residual energy density at the end of the quench is simply proportional to the defect density. Hence, all of our results for the defect density can be recontextualized in terms of the residual energy density.]
Unlike a zero-noise system in which $\varepsilon_\mathrm{exc}=0$, our quantum simulations shown in Fig.~\ref{fig:Anti-KZ_ISING}(d) reveal an excess energy that grows with quench time and the corresponding increased circuit depth.
In particular, over a range of system sizes, $N=80$, $100$, and $120$ qubits, we find evidence for the power-law scaling relation $\varepsilon_\mathrm{exc}(\tau_Q) \propto \tau_Q^{-\gamma}$ with $\gamma\approx -0.6$.
Such scaling can only emerge due to the noisy environment of the quantum simulator, and we speculate that its consistent value across various system sizes reveals a universal character about the nature of interactions between the transmon qubits and their environment.

\vspace{0.5cm}
{\textbf{Discussion and outlook}}

In this paper, we have demonstrated that noise can fundamentally alter universal scaling as exhibited through connected correlations, defect densities, and excess energies, shedding light on emergent scaling behavior in noisy environments. Our results complement and extend the conclusions of~\cite{Miessen2024} by demonstrating that signatures of universal scaling can be robustly observed even when the microscopic noise model of the hardware is unknown, provided that the device operates in a sufficiently low cross-talk and low-error regime.
While~\cite{Miessen2024} emphasizes the role of hardware improvements (in particular, reduced cross-talk and faster entangling gates) in extending the temporal window over which QKZ scaling is visible, our experiments show that on newer-generation hardware signatures of universal scaling behavior can persist across multiple system sizes and quench protocols without the need for explicit noise-mitigation techniques.

In contrast to earlier observations where noise-induced defect production masked universal behavior at long times, our measurements retain a clear power-law scaling across a broad range of quench times, indicating emergent universality in the many-body dynamics in the presence of noise.
Additionally, our work advances the use of QKZ dynamics from a diagnostic of \emph{when} universal scaling breaks down to a tool for \emph{what} scaling is realized.
By simultaneously extracting scaling exponents for both the defect density and the excess energy, we are able to distinguish deviations from the closed-system universal dynamics in a controlled and quantitative manner.
This suggests that universal nonequilibrium scaling may serve not only as a pass-fail benchmark, but as a framework for characterizing effective universality classes associated with different quantum hardware platforms.
In this perspective, distinct noise structures, such as dephasing-dominated versus cross-talk-dominated errors, may renormalize the effective critical dynamics in different ways, leading to measurable shifts in the observed scaling exponents.
Our results therefore point toward the possibility of using universal dynamical scaling as a high-level descriptor of quantum hardware, complementary to conventional gate-level performance metrics.

Finally, while the present study focuses on one-dimensional dynamics, the approach naturally extends to two-dimensional systems, where classical simulation methods rapidly become intractable and where noise-induced effects are expected to interplay more strongly with critical dynamics.
Extending QKZ protocols to two-dimensional lattices would enable the exploration of geometry-dependent light-cones, defect coarsening, and universality classes that are fundamentally inaccessible in one dimension.
Such experiments would not only provide a stringent test of quantum advantage in nonequilibrium quantum dynamics but also open a path toward using universal scaling laws as a unifying language for comparing quantum simulators across architectures and spatial dimensions.

\vspace{0.5cm}
{\textbf{Acknowledgements.}}
The authors would like to thank Oles Shtanko and Yi-Zhuang You for insightful discussions.
B.R. gratefully acknowledges funding from the DLR Space Administration with funds provided by the Federal Ministry for Economic Affairs and Climate Action (BMWK) under grant numbers 50WM2451 (QUANTUMANIA) and from the Deutsche Forschungsgemeinschaft (DFG) through SFB 1227 (DQ-mat) within Project A05, Germany’s Excellence Strategy (EXC-2123 QuantumFrontiers Grants No. 390837967).
B.R. acknowledges the support of the Quantum Leap Challenge Institute for Hybrid Quantum Architectures and Networks Grant No. OMA2016136.
This research is funded in part by a QuantEmX grant from ICAM and the Gordon and Betty Moore Foundation through Grant GBMF9616 to Brendan Rhyno.
S.V. acknowledges the support of the National Aeronautics and Space Administration (NASA) Science Mission Directorate, Division of Biological and Physical Sciences (BPS) through ROSES-22 as well as multiple Jet Propulsion Laboratory (JPL) Research Support Agreements.
We acknowledge the use of IBM Quantum Credits for this work. The views expressed are those of the authors, and do not reflect the official policy or position of IBM or the IBM Quantum team.

\let\oldaddcontentsline\addcontentsline
\renewcommand{\addcontentsline}[3]{}
\bibliography{references}

@book{Sachdev2001,
  author        = {Sachdev, Subir},
  title         = {{Quantum Phase Transitions}},
  year          = {2001},
  publisher     = {Cambridge University Press}
}

@book{Lewenstein2012,
  author        = {Maciej Lewenstein and Anna Sanpera and Veronica Ahufinger},
  title         = {{Ultracold Atoms in Optical Lattices: Simulating Quantum Many-Body Systems}},
  publisher     = {OUP Oxford},
  year          = {2012},
  address       = {Oxford},
  isbn          = {9780199573127}
}

@book{Nielsen2001,
  author        = {Nielsen, Michael A and Chuang, Isaac L},
  title         = {{Quantum Computation and Quantum Information}},
  year          = {2001},
  publisher     = {Cambridge University Press}
}

@article{Zurek2005,
  author        = {Zurek, W. H. and Dorner, U. and Zoller, P.},
  title         = {{Dynamics of a quantum phase transition}},
  journal       = {Physical Review Letters},
  volume        = {95},
  pages         = {105701},
  year          = {2005},
  doi           = {10.1103/PhysRevLett.95.105701}
}

@article{Dziarmaga2010,
  title         = {{Dynamics of a quantum phase transition and relaxation to a steady state}},
  author        = {Dziarmaga, Jacek},
  journal       = {Advances in Physics},
  volume        = {59},
  number        = {6},
  pages         = {1063--1189},
  year          = {2010},
  publisher     = {Taylor \& Francis},
  doi           = {10.1080/00018732.2010.514702},
  url           = {https://doi.org/10.1080/00018732.2010.514702}
}

@article{Polkovnikov2005,
  author        = {Polkovnikov},
  title         = {{Universal adiabatic dynamics in the vicinity of a quantum critical point}},
  journal       = {Physical Review B},
  volume        = {72},
  pages         = {161201},
  year          = {2005},
  doi           = {10.1103/PhysRevB.72.161201}
}

@article{Cirac2012,
  author        = {J. I. Cirac and P. Zoller},
  title         = {{Goals and opportunities in quantum simulation}},
  journal       = {Nature Physics},
  volume        = {8},
  pages         = {264--266},
  year          = {2012},
  doi           = {10.1038/nphys2275}
}

@article{Daley2022,
  author        = {Daley, Andrew J. and Bloch, Immanuel and Kokail, Christian and Flannigan, Stuart and Pearson, Natalie and Troyer, Matthias and Zoller, Peter},
  title         = {{Practical quantum advantage in quantum simulation}},
  journal       = {Nature},
  volume        = {607},
  issue         = {7920},
  pages         = {667--676},
  year          = {2022},
  doi           = {10.1038/s41586-022-04940-6},
  url           = {https://doi.org/10.1038/s41586-022-04940-6}
}

@article{Bloch2008,
  author        = {Bloch, I. and Dalibard, J. and Zwerger, W.},
  title         = {{Many-body physics with ultracold gases}},
  journal       = {Reviews of Modern Physics},
  volume        = {80},
  pages         = {885--964},
  year          = {2008},
  doi           = {10.1103/RevModPhys.80.885}
}

@article{Blatt2012,
  author        = {Blatt, R. and Roos, C. F.},
  title         = {{Quantum simulations with trapped ions}},
  journal       = {Nature Physics},
  volume        = {8},
  pages         = {277--284},
  year          = {2012},
  doi           = {10.1038/nphys2252}
}

@article{Houck2012,
  author        = {Houck, A. A. and Türeci, H. E. and Koch, J.},
  title         = {{On-chip quantum simulation with superconducting circuits}},
  journal       = {Nature Physics},
  volume        = {8},
  pages         = {292--299},
  year          = {2012},
  doi           = {10.1038/nphys2251}
}

@article{Browaeys2020,
  author        = {Browaeys, A. and Lahaye, T.},
  title         = {{Many-body physics with individually controlled Rydberg atoms}},
  journal       = {Nature Physics},
  volume        = {16},
  pages         = {132--142},
  year          = {2020},
  doi           = {10.1038/s41567-019-0733-z}
}

@article{Monroe2021,
  author        = {Christopher Monroe and W. C. Campbell and L.-M. Duan and Z.-X. Gong and A. V. Gorshkov and P. W. Hess and R. Islam and Kihwan Kim and N. M. Linke and G. Pagano and P. Richerme and C. Senko and Norman Y. Yao},
  title         = {{Programmable Quantum Simulations of Spin Systems with Trapped Ions}},
  journal       = {Reviews of Modern Physics},
  volume        = {93},
  number        = {2},
  pages         = {025001},
  year          = {2021},
  month         = {April},
  doi           = {10.1103/RevModPhys.93.025001},
  url           = {https://link.aps.org/doi/10.1103/RevModPhys.93.025001}
}

@article{Koyluouglu2026,
  title         = {{Measuring central charge on a universal quantum processor}},
  author        = {K{\"o}yl{\"u}o{\u{g}}lu, Nazl{\i} U{\u{g}}ur and Majumder, Swarnadeep and Amico, Mirko and Mostame, Sarah and van den Berg, Ewout and Rajabpour, M. A. and Minev, Zlatko and Najafi, Khadijeh},
  journal       = {Nature Communications},
  volume        = {17},
  issue         = {1},
  pages         = {305},
  year          = {2026},
  publisher     = {Nature Publishing Group},
  doi           = {10.1038/s41467-025-66775-9},
  url           = {https://doi.org/10.1038/s41467-025-66775-9}
}

@article{Miessen2024,
  author        = {Miessen, Alexander and Egger, Daniel J. and Tavernelli, Ivano and Mazzola, Guglielmo},
  title         = {{Benchmarking Digital Quantum Simulations Above Hundreds of Qubits Using Quantum Critical Dynamics}},
  journal       = {PRX Quantum},
  volume        = {5},
  number        = {4},
  pages         = {040320},
  year          = {2024},
  doi           = {10.1103/PRXQuantum.5.040320},
  url           = {https://doi.org/10.1103/PRXQuantum.5.040320}
}

@article{Keesling2019,
  author        = {Keesling, Alexander and Omran, Ahmed and Levine, Harry and Bernien, Hannes and Pichler, Hannes and Choi, Soonwon and Samajdar, Rhine and Schwartz, Sylvain and Silvi, Pietro and Sachdev, Subir and Zoller, Peter and Endres, Manuel and Greiner, Markus and Vuleti{\'c}, Vladan and Lukin, Mikhail D.},
  year          = {2019},
  title         = {{Quantum Kibble–Zurek mechanism and critical dynamics on a programmable Rydberg simulator}},
  journal       = {Nature},
  volume        = {568},
  issue         = {7751},
  pages         = {207--211},
  month         = {Apr},
  publisher     = {Nature Publishing Group},
  doi           = {10.1038/s41586-019-1070-1},
  url           = {https://doi.org/10.1038/s41586-019-1070-1}
}

@article{Kato1950,
  title         = {{On the adiabatic theorem of quantum mechanics}},
  author        = {Kato, Tosio},
  journal       = {Journal of the Physical Society of Japan},
  volume        = {5},
  number        = {6},
  pages         = {435--439},
  year          = {1950},
  publisher     = {The Physical Society of Japan},
  doi           = {10.1143/JPSJ.5.435},
  url           = {https://doi.org/10.1143/JPSJ.5.435}
}

@article{Suzuki2007,
  author        = {S. Suzuki and H. Nishimori and M. Suzuki},
  title         = {{Quantum annealing of the random-field Ising model by transverse ferromagnetic interactions}},
  journal       = {Physical Review E},
  volume        = {75},
  pages         = {051112},
  year          = {2007},
  doi           = {10.1103/PhysRevE.75.051112}
}

@article{Avron1999,
  author        = {J. E. Avron and A. Elgart},
  title         = {{Adiabatic theorem without a gap condition}},
  journal       = {Communications in Mathematical Physics},
  volume        = {203},
  pages         = {445--463},
  year          = {1999},
  doi           = {10.1007/s002200050614}
}

@article{Francuz2016,
  title         = {{Space and time renormalization in phase transition dynamics}},
  author        = {Francuz, Anna and Dziarmaga, Jacek and Gardas, Bart\l{}omiej and Zurek, Wojciech H.},
  journal       = {Phys. Rev. B},
  volume        = {93},
  issue         = {7},
  pages         = {075134},
  numpages      = {14},
  year          = {2016},
  month         = {Feb},
  publisher     = {American Physical Society},
  doi           = {10.1103/PhysRevB.93.075134},
  url           = {https://link.aps.org/doi/10.1103/PhysRevB.93.075134}
}

@article{Nalbach2015,
  title         = {{Quantum Kibble-Zurek physics in the presence of spatially correlated dissipation}},
  author        = {Nalbach, P. and Vishveshwara, Smitha and Clerk, Aashish A.},
  journal       = {Phys. Rev. B},
  volume        = {92},
  issue         = {1},
  pages         = {014306},
  numpages      = {8},
  year          = {2015},
  month         = {Jul},
  publisher     = {American Physical Society},
  doi           = {10.1103/PhysRevB.92.014306},
  url           = {https://link.aps.org/doi/10.1103/PhysRevB.92.014306}
}

@article{Zalka1998,
  title         = {{Simulating quantum systems on a quantum computer}},
  author        = {Zalka, Christof},
  journal       = {Proceedings of the Royal Society of London. Series A: Mathematical, Physical and Engineering Sciences},
  volume        = {454},
  number        = {1969},
  pages         = {313-322},
  year          = {1998},
  doi           = {10.1098/rspa.1998.0162},
  url           = {https://royalsocietypublishing.org/doi/abs/10.1098/rspa.1998.0162},
  publisher     = {The Royal Society}
}

@article{Miessen2023,
  author        = {Miessen, Alexander and Ollitrault, Pauline J. and Tacchino, Francesco and Tavernelli, Ivano},
  year          = {2023},
  title         = {{Quantum algorithms for quantum dynamics}},
  journal       = {Nature Computational Science},
  volume        = {3},
  issue         = {1},
  pages         = {25--37},
  month         = {Jan},
  publisher     = {Nature Publishing Group},
  doi           = {10.1038/s43588-022-00374-2},
  url           = {https://doi.org/10.1038/s43588-022-00374-2}
}

@article{Dutta2016,
  title         = {{Anti-Kibble-Zurek Behavior in Crossing the Quantum Critical Point of a Thermally Isolated System Driven by a Noisy Control Field}},
  author        = {Anirban Dutta and Armin Rahmani and Adolfo del Campo},
  journal       = {Physical Review Letters},
  volume        = {117},
  number        = {8},
  pages         = {080402},
  year          = {2016},
  doi           = {10.1103/PhysRevLett.117.080402},
  url           = {https://journals.aps.org/prl/abstract/10.1103/PhysRevLett.117.080402}
}

@article{Keck2017,
  doi           = {10.1088/1367-2630/aa8cef},
  url           = {https://dx.doi.org/10.1088/1367-2630/aa8cef},
  year          = {2017},
  publisher     = {IOP Publishing},
  volume        = {19},
  number        = {11},
  pages         = {113029},
  author        = {Maximilian Keck and Simone Montangero and Giuseppe E Santoro and Rosario Fazio and Davide Rossini},
  title         = {{Dissipation in adiabatic quantum computers: lessons from an exactly solvable model}},
  journal       = {New Journal of Physics}
}

@article{Kuo2021,
  title         = {{Decoherent quench dynamics across quantum phase transitions}},
  author        = {Wei-Ting Kuo and Daniel Arovas and Smitha Vishveshwara and Yi-Zhuang You},
  journal       = {SciPost Phys.},
  volume        = {11},
  pages         = {084},
  year          = {2021},
  publisher     = {SciPost},
  doi           = {10.21468/SciPostPhys.11.4.084},
  url           = {https://scipost.org/10.21468/SciPostPhys.11.4.084}
}

@article{Kibble1976,
  title         = {{Topology of cosmic domains and strings}},
  author        = {Kibble, T. W. B.},
  year          = {1976},
  publisher     = {{IOP} Publishing},
  volume        = {9},
  issue         = {8},
  pages         = {1387--1398},
  journal       = {Journal of Physics A: Mathematical and General},
  doi           = {10.1088/0305-4470/9/8/029},
  url           = {https://doi.org/10.1088/0305-4470/9/8/029}
}

@article{Zurek1985,
  title         = {{Cosmological experiments in superfluid helium?}},
  author        = {Zurek, W. H.},
  journal       = {Nature},
  volume        = {317},
  issue         = {6037},
  pages         = {505--508},
  year          = {1985},
  publisher     = {Nature Publishing Group},
  doi           = {10.1038/317505a0},
  url           = {https://doi.org/10.1038/317505a0}
}

@article{Zurek1996,
  title         = {{Cosmological experiments in condensed matter systems}},
  author        = {Zurek, W. H.},
  journal       = {Physics Reports},
  volume        = {276},
  issue         = {4},
  pages         = {177--221},
  year          = {1996},
  publisher     = {Elsevier},
  doi           = {10.1016/S0370-1573(96)00009-9},
  url           = {https://doi.org/10.1016/S0370-1573(96)00009-9}
}

@article{Kibble1980,
  title         = {{Some implications of a cosmological phase transition}},
  author        = {Kibble, T. W. B.},
  journal       = {Physics Reports},
  volume        = {67},
  issue         = {1},
  pages         = {183--199},
  year          = {1980},
  publisher     = {Elsevier},
  doi           = {10.1016/0370-1573(80)90091-5},
  url           = {https://doi.org/10.1016/0370-1573(80)90091-5},
}

@article{Dziarmaga2005,
  title         = {{Dynamics of a Quantum Phase Transition: Exact Solution of the Quantum Ising Model}},
  author        = {Dziarmaga, Jacek},
  journal       = {Phys. Rev. Lett.},
  volume        = {95},
  issue         = {24},
  pages         = {245701},
  numpages      = {4},
  year          = {2005},
  publisher     = {American Physical Society},
  doi           = {10.1103/PhysRevLett.95.245701},
  url           = {https://link.aps.org/doi/10.1103/PhysRevLett.95.245701}
}

@article{Polkovnikov2011,
  title         = {{Colloquium: Nonequilibrium dynamics of closed interacting quantum systems}},
  author        = {Polkovnikov, Anatoli and Sengupta, Krishnendu and Silva, Alessandro and Vengalattore, Mukund},
  journal       = {Rev. Mod. Phys.},
  volume        = {83},
  issue         = {3},
  pages         = {863--883},
  numpages      = {0},
  year          = {2011},
  month         = {Aug},
  publisher     = {American Physical Society},
  doi           = {10.1103/RevModPhys.83.863},
  url           = {https://link.aps.org/doi/10.1103/RevModPhys.83.863}
}

@article{Smith2019,
  author        = {Smith, Adam and Kim, M. S. and Pollmann, Frank and Knolle, Johannes},
  year          = {2019},
  title         = {{Simulating quantum many-body dynamics on a current digital quantum computer}},
  journal       = {npj Quantum Information},
  volume        = {5},
  issue         = {1},
  pages         = {106},
  month         = {Nov},
  publisher     = {Nature Publishing Group},
  doi           = {10.1038/s41534-019-0217-0},
  url           = {https://doi.org/10.1038/s41534-019-0217-0}
}

@article{Dupont2022,
  title         = {{Quantum criticality using a superconducting quantum processor}},
  author        = {Dupont, Maxime and Moore, Joel E.},
  journal       = {Phys. Rev. B},
  volume        = {106},
  issue         = {4},
  pages         = {L041109},
  numpages      = {7},
  year          = {2022},
  month         = {Jul},
  publisher     = {American Physical Society},
  doi           = {10.1103/PhysRevB.106.L041109},
  url           = {https://link.aps.org/doi/10.1103/PhysRevB.106.L041109}
}

@article{Poulin2011,
  title         = {{Quantum Simulation of Time-Dependent Hamiltonians and the Convenient Illusion of Hilbert Space}},
  author        = {Poulin, David and Qarry, Angie and Somma, Rolando and Verstraete, Frank},
  journal       = {Phys. Rev. Lett.},
  volume        = {106},
  issue         = {17},
  pages         = {170501},
  numpages      = {4},
  year          = {2011},
  month         = {Apr},
  publisher     = {American Physical Society},
  doi           = {10.1103/PhysRevLett.106.170501},
  url           = {https://link.aps.org/doi/10.1103/PhysRevLett.106.170501}
}

@article{Trotter1959,
  author        = {H. F. Trotter},
  title         = {{On the Product of Semi-Groups of Operators}},
  url           = {http://www.jstor.org/stable/2033649},
  journal       = {Proceedings of the American Mathematical Society},
  number        = {4},
  pages         = {545--551},
  publisher     = {American Mathematical Society},
  volume        = {10},
  year          = {1959}
}

@article{Suzuki1993,
  title         = {{General Decomposition Theory of Ordered Exponentials}},
  author        = {Suzuki, Masuo},
  journal       = {Proceedings of the Japan Academy, Series B},
  volume        = {69},
  number        = {7},
  pages         = {161-166},
  year          = {1993},
  publisher     = {The Japan Academy},
  doi           = {10.2183/pjab.69.161},
  url           = {https://doi.org/10.2183/pjab.69.161}
}

@article{Lloyd1996,
  author        = {Seth Lloyd},
  title         = {{Universal Quantum Simulators}},
  journal       = {Science},
  volume        = {273},
  number        = {5278},
  pages         = {1073-1078},
  year          = {1996},
  doi           = {10.1126/science.273.5278.1073},
  url           = {https://www.science.org/doi/abs/10.1126/science.273.5278.1073}
}

@article{Lieb1961,
  title         = {{Two soluble models of an antiferromagnetic chain}},
  author        = {Lieb, Elliott and Schultz, Theodore and Mattis, Daniel},
  journal       = {Annals of Physics},
  volume        = {16},
  number        = {3},
  pages         = {407--466},
  year          = {1961},
  publisher     = {Elsevier},
  doi           = {https://doi.org/10.1016/0003-4916(61)90115-4},
  url           = {https://www.sciencedirect.com/science/article/pii/0003491661901154}
}

@article{Barouch1971,
  title         = {{Statistical Mechanics of the $XY$ Model. II. Spin-Correlation Functions}},
  author        = {Barouch, Eytan and McCoy, Barry M.},
  journal       = {Phys. Rev. A},
  volume        = {3},
  issue         = {2},
  pages         = {786--804},
  year          = {1971},
  month         = {Feb},
  publisher     = {American Physical Society},
  doi           = {10.1103/PhysRevA.3.786},
  url           = {https://link.aps.org/doi/10.1103/PhysRevA.3.786}
}

@article{McCoy1971,
  title         = {{Statistical Mechanics of the $\mathrm{XY}$ Model. IV. Time-Dependent Spin-Correlation Functions}},
  author        = {McCoy, Barry M. and Barouch, Eytan and Abraham, Douglas B.},
  journal       = {Phys. Rev. A},
  volume        = {4},
  issue         = {6},
  pages         = {2331--2341},
  year          = {1971},
  month         = {Dec},
  publisher     = {American Physical Society},
  doi           = {10.1103/PhysRevA.4.2331},
  url           = {https://link.aps.org/doi/10.1103/PhysRevA.4.2331}
}

@article{Wimmer2012,
  title         = {{Efficient numerical computation of the pfaffian for dense and banded skew-symmetric matrices}},
  author        = {Michael Wimmer},
  journal       = {ACM Transactions on Mathematical Software (TOMS)},
  volume        = {38},
  number        = {4},
  pages         = {1--17},
  year          = {2012},
  publisher     = {ACM New York, NY, USA}
}

@article{Gao2017,
  title         = {{Anti-Kibble-Zurek behavior of a noisy transverse-field $\mathrm{XY}$ chain and its quantum simulation with two-level systems}},
  author        = {Gao, Zhi-Peng and Zhang, Dan-Wei and Yu, Yang and Zhu, Shi-Liang},
  journal       = {Phys. Rev. B},
  volume        = {95},
  issue         = {22},
  pages         = {224303},
  numpages      = {8},
  year          = {2017},
  month         = {Jun},
  publisher     = {American Physical Society},
  doi           = {10.1103/PhysRevB.95.224303},
  url           = {https://link.aps.org/doi/10.1103/PhysRevB.95.224303}
}

@article{Gangadharaiah2021,
  title         = {{Driven quantum spin chain in the presence of noise: Anti-Kibble-Zurek behavior}},
  author        = {Singh, Manvendra and Gangadharaiah, Suhas},
  journal       = {Phys. Rev. B},
  volume        = {104},
  issue         = {6},
  pages         = {064313},
  numpages      = {11},
  year          = {2021},
  month         = {Aug},
  publisher     = {American Physical Society},
  doi           = {10.1103/PhysRevB.104.064313},
  url           = {https://link.aps.org/doi/10.1103/PhysRevB.104.064313}
}

@article{SciPy2020,
  author        = {Virtanen, Pauli and Gommers, Ralf and Oliphant, Travis E. and Haberland, Matt and Reddy, Tyler and Cournapeau, David and Burovski, Evgeni and Peterson, Pearu and Weckesser, Warren and Bright, Jonathan and {van der Walt}, St{\'e}fan J. and Brett, Matthew and Wilson, Joshua and Millman, K. Jarrod and Mayorov, Nikolay and Nelson, Andrew R. J. and Jones, Eric and Kern, Robert and Larson, Eric and Carey, C J and Polat, {\.I}lhan and Feng, Yu and Moore, Eric W. and {VanderPlas}, Jake and Laxalde, Denis and Perktold, Josef and Cimrman, Robert and Henriksen, Ian and Quintero, E. A. and Harris, Charles R. and Archibald, Anne M. and Ribeiro, Ant{\^o}nio H. and Pedregosa, Fabian and {van Mulbregt}, Paul and {SciPy 1.0 Contributors}},
  title         = {{{SciPy} 1.0: Fundamental Algorithms for Scientific
            Computing in Python}},
  journal       = {Nature Methods},
  year          = {2020},
  volume        = {17},
  pages         = {261--272},
  adsurl        = {https://rdcu.be/b08Wh},
  doi           = {10.1038/s41592-019-0686-2},
}

@article{Andersen2025,
  title         = {{Thermalization and criticality on an analogue–digital quantum simulator}},
  author        = {Andersen, T. I. and Astrakhantsev, N. and Karamlou, A. H. and Berndtsson, J. and Motruk, J. and Szasz, A. and Gross, J. A. and Schuckert, A. and Westerhout, T. and Zhang, Y. and Forati, E. and Rossi, D. and Kobrin, B. and Paolo, A. Di and Klots, A. R. and Drozdov, I. and Kurilovich, V. and Petukhov, A. and Ioffe, L. B. and Elben, A. and Rath, A. and Vitale, V. and Vermersch, B. and Acharya, R. and Beni, L. A. and Anderson, K. and Ansmann, M. and Arute, F. and Arya, K. and Asfaw, A. and Atalaya, J. and Ballard, B. and Bardin, J. C. and Bengtsson, A. and Bilmes, A. and Bortoli, G. and Bourassa, A. and Bovaird, J. and Brill, L. and Broughton, M. and Browne, D. A. and Buchea, B. and Buckley, B. B. and Buell, D. A. and Burger, T. and Burkett, B. and Bushnell, N. and Cabrera, A. and Campero, J. and Chang, H.-S. and Chen, Z. and Chiaro, B. and Claes, J. and Cleland, A. Y. and Cogan, J. and Collins, R. and Conner, P. and Courtney, W. and Crook, A. L. and Das, S. and Debroy, D. M. and Lorenzo, L. De and Barba, A. Del Toro and Demura, S. and Donohoe, P. and Dunsworth, A. and Earle, C. and Eickbusch, A. and Elbag, A. M. and Elzouka, M. and Erickson, C. and Faoro, L. and Fatemi, R. and Ferreira, V. S. and Burgos, L. Flores and Fowler, A. G. and Foxen, B. and Ganjam, S. and Gasca, R. and Giang, W. and Gidney, C. and Gilboa, D. and Giustina, M. and Gosula, R. and Dau, A. Grajales and Graumann, D. and Greene, A. and Habegger, S. and Hamilton, M. C. and Hansen, M. and Harrigan, M. P. and Harrington, S. D. and Heslin, S. and Heu, P. and Hill, G. and Hoffmann, M. R. and Huang, H.-Y. and Huang, T. and Huff, A. and Huggins, W. J. and Isakov, S. V. and Jeffrey, E. and Jiang, Z. and Jones, C. and Jordan, S. and Joshi, C. and Juhas, P. and Kafri, D. and Kang, H. and Kechedzhi, K. and Khaire, T. and Khattar, T. and Khezri, M. and Kieferová, M. and Kim, S. and Kitaev, A. and Klimov, P. and Korotkov, A. N. and Kostritsa, F. and Kreikebaum, J. M. and Landhuis, D. and Langley, B. W. and Laptev, P. and Lau, K.-M. and Guevel, L. Le and Ledford, J. and Lee, J. and Lee, K. W. and Lensky, Y. D. and Lester, B. J. and Li, W. Y. and Lill, A. T. and Liu, W. and Livingston, W. P. and Locharla, A. and Lundahl, D. and Lunt, A. and Madhuk, S. and Maloney, A. and Mandrà, S. and Martin, L. S. and Martin, O. and Martin, S. and Maxfield, C. and McClean, J. R. and McEwen, M. and Meeks, S. and Miao, K. C. and Mieszala, A. and Molina, S. and Montazeri, S. and Morvan, A. and Movassagh, R. and Neill, C. and Nersisyan, A. and Newman, M. and Nguyen, A. and Nguyen, M. and Ni, C.-H. and Niu, M. Y. and Oliver, W. D. and Ottosson, K. and Pizzuto, A. and Potter, R. and Pritchard, O. and Pryadko, L. P. and Quintana, C. and Reagor, M. J. and Rhodes, D. M. and Roberts, G. and Rocque, C. and Rosenberg, E. and Rubin, N. C. and Saei, N. and Sankaragomathi, K. and Satzinger, K. J. and Schurkus, H. F. and Schuster, C. and Shearn, M. J. and Shorter, A. and Shutty, N. and Shvarts, V. and Sivak, V. and Skruzny, J. and Small, S. and Smith, W. Clarke and Springer, S. and Sterling, G. and Suchard, J. and Szalay, M. and Sztein, A. and Thor, D. and Torres, A. and Torunbalci, M. M. and Vaishnav, A. and Vdovichev, S. and Villalonga, B. and Heidweiller, C. Vollgraff and Waltman, S. and Wang, S. X. and White, T. and Wong, K. and Woo, B. W. K. and Xing, C. and Yao, Z. Jamie and Yeh, P. and Ying, B. and Yoo, J. and Yosri, N. and Young, G. and Zalcman, A. and Zhu, N. and Zobrist, N. and Neven, H. and Babbush, R. and Boixo, S. and Hilton, J. and Lucero, E. and Megrant, A. and Kelly, J. and Chen, Y. and Smelyanskiy, V. and Vidal, G. and Roushan, P. and Läuchli, A. M. and Abanin, D. A. and Mi, X.},
  journal       = {Nature},
  publisher     = {Nature Publishing Group},
  year          = {2025},
  volume        = {638},
  issue         = {8049},
  pages         = {79--85},
  doi           = {10.1038/s41586-024-08460-3},
  url           = {https://doi.org/10.1038/s41586-024-08460-3}
}

@article{Manovitz2025,
  title         = {{Quantum coarsening and collective dynamics on a programmable simulator}},
  author        = {Manovitz, Tom and Li, Sophie H. and Ebadi, Sepehr and Samajdar, Rhine and Geim, Alexandra A. and Evered, Simon J. and Bluvstein, Dolev and Zhou, Hengyun and Koyluoglu, Nazlı Uğur and Feldmeier, Johannes and Dolgirev, Pavel E. and Maskara, Nishad and Kalinowski, Marcin and Sachdev, Subir and Huse, David A. and Greiner, Markus and Vuletić, Vladan and Lukin, Mikhail D.},
  journal       = {Nature},
  year          = {2025},
  volume        = {638},
  number        = {8049},
  pages         = {86--92},
  doi           = {10.1038/s41586-024-08353-5}
}

@article{Minev2025,
  title         = {{Realizing string-net condensation: Fibonacci anyon braiding for universal gates and sampling chromatic polynomials}},
  author        = {Minev, Zlatko K. and Najafi, Khadijeh and Majumder, Swarnadeep and Wang, Juven and Stern, Ady and Kim, Eun-Ah and Jian, Chao-Ming and Zhu, Guanyu},
  journal       = {Nature Communications},
  volume        = {16},
  issue         = {1},
  pages         = {6225},
  year          = {2025},
  publisher     = {Nature Publishing Group},
  doi           = {10.1038/s41467-025-61493-8},
  url           = {https://doi.org/10.1038/s41467-025-61493-8}
}

@article{King2025,
  author        = {Andrew D. King  and Alberto Nocera  and Marek M. Rams  and Jacek Dziarmaga  and Roeland Wiersema  and William Bernoudy  and Jack Raymond  and Nitin Kaushal  and Niclas Heinsdorf  and Richard Harris  and Kelly Boothby  and Fabio Altomare  and Mohsen Asad  and Andrew J. Berkley  and Martin Boschnak  and Kevin Chern  and Holly Christiani  and Samantha Cibere  and Jake Connor  and Martin H. Dehn  and Rahul Deshpande  and Sara Ejtemaee  and Pau Farre  and Kelsey Hamer  and Emile Hoskinson  and Shuiyuan Huang  and Mark W. Johnson  and Samuel Kortas  and Eric Ladizinsky  and Trevor Lanting  and Tony Lai  and Ryan Li  and Allison J. R. MacDonald  and Gaelen Marsden  and Catherine C. McGeoch  and Reza Molavi  and Travis Oh  and Richard Neufeld  and Mana Norouzpour  and Joel Pasvolsky  and Patrick Poitras  and Gabriel Poulin-Lamarre  and Thomas Prescott  and Mauricio Reis  and Chris Rich  and Mohammad Samani  and Benjamin Sheldan  and Anatoly Smirnov  and Edward Sterpka  and Berta Trullas Clavera  and Nicholas Tsai  and Mark Volkmann  and Alexander M. Whiticar  and Jed D. Whittaker  and Warren Wilkinson  and Jason Yao  and T. J. Yi  and Anders W. Sandvik  and Gonzalo Alvarez  and Roger G. Melko  and Juan Carrasquilla  and Marcel Franz  and Mohammad H. Amin},
  title         = {{Beyond-classical computation in quantum simulation}},
  journal       = {Science},
  volume        = {388},
  number        = {6743},
  pages         = {199-204},
  year          = {2025},
  doi           = {10.1126/science.ado6285},
  url           = {https://www.science.org/doi/abs/10.1126/science.ado6285}
}

@article{Schuetzhold2006,
  title         = {{Sweeping from the Superfluid to the Mott Phase in the Bose-Hubbard Model}},
  author        = {Sch\"utzhold, Ralf and Uhlmann, Michael and Xu, Yan and Fischer, Uwe R.},
  journal       = {Phys. Rev. Lett.},
  volume        = {97},
  issue         = {20},
  pages         = {200601},
  numpages      = {4},
  year          = {2006},
  publisher     = {American Physical Society},
  doi           = {10.1103/PhysRevLett.97.200601},
  url           = {https://link.aps.org/doi/10.1103/PhysRevLett.97.200601}
}

@article{Uhlmann2007,
  title         = {{Vortex Quantum Creation and Winding Number Scaling in a Quenched Spinor Bose Gas}},
  author        = {Uhlmann, Michael and Sch\"utzhold, Ralf and Fischer, Uwe R.},
  journal       = {Phys. Rev. Lett.},
  volume        = {99},
  issue         = {12},
  pages         = {120407},
  numpages      = {4},
  year          = {2007},
  publisher     = {American Physical Society},
  doi           = {10.1103/PhysRevLett.99.120407},
  url           = {https://link.aps.org/doi/10.1103/PhysRevLett.99.120407}
}

@article{Uhlmann2010,
  title         = {{$O(N)$ symmetry-breaking quantum quench: Topological defects versus quasiparticles}},
  author        = {Uhlmann, Michael and Sch\"utzhold, Ralf and Fischer, Uwe R.},
  journal       = {Phys. Rev. D},
  volume        = {81},
  issue         = {2},
  pages         = {025017},
  numpages      = {6},
  year          = {2010},
  publisher     = {American Physical Society},
  doi           = {10.1103/PhysRevD.81.025017},
  url           = {https://link.aps.org/doi/10.1103/PhysRevD.81.025017}
}

@article{Fauseweh2024,
  title         = {{Quantum many-body simulations on digital quantum computers: State-of-the-art and future challenges}},
  author        = {Fauseweh, Benedikt},
  journal       = {Nature Communications},
  volume        = {15},
  issue         = {1},
  pages         = {2123},
  year          = {2024},
  publisher     = {Nature Publishing Group},
  doi           = {10.1038/s41467-024-46402-9},
  url           = {https://doi.org/10.1038/s41467-024-46402-9}
}

@article{Mi2022,
  author        = {X. Mi  and M. Sonner  and M. Y. Niu  and K. W. Lee  and B. Foxen  and R. Acharya  and I. Aleiner  and T. I. Andersen  and F. Arute  and K. Arya  and A. Asfaw  and J. Atalaya  and J. C. Bardin  and J. Basso  and A. Bengtsson  and G. Bortoli  and A. Bourassa  and L. Brill  and M. Broughton  and B. B. Buckley  and D. A. Buell  and B. Burkett  and N. Bushnell  and Z. Chen  and B. Chiaro  and R. Collins  and P. Conner  and W. Courtney  and A. L. Crook  and D. M. Debroy  and S. Demura  and A. Dunsworth  and D. Eppens  and C. Erickson  and L. Faoro  and E. Farhi  and R. Fatemi  and L. Flores  and E. Forati  and A. G. Fowler  and W. Giang  and C. Gidney  and D. Gilboa  and M. Giustina  and A. G. Dau  and J. A. Gross  and S. Habegger  and M. P. Harrigan  and M. Hoffmann  and S. Hong  and T. Huang  and A. Huff  and W. J. Huggins  and L. B. Ioffe  and S. V. Isakov  and J. Iveland  and E. Jeffrey  and Z. Jiang  and C. Jones  and D. Kafri  and K. Kechedzhi  and T. Khattar  and S. Kim  and A. Y. Kitaev  and P. V. Klimov  and A. R. Klots  and A. N. Korotkov  and F. Kostritsa  and J. M. Kreikebaum  and D. Landhuis  and P. Laptev  and K.-M. Lau  and J. Lee  and L. Laws  and W. Liu  and A. Locharla  and O. Martin  and J. R. McClean  and M. McEwen  and B. Meurer Costa  and K. C. Miao  and M. Mohseni  and S. Montazeri  and A. Morvan  and E. Mount  and W. Mruczkiewicz  and O. Naaman  and M. Neeley  and C. Neill  and M. Newman  and T. E. O’Brien  and A. Opremcak  and A. Petukhov  and R. Potter  and C. Quintana  and N. C. Rubin  and N. Saei  and D. Sank  and K. Sankaragomathi  and K. J. Satzinger  and C. Schuster  and M. J. Shearn  and V. Shvarts  and D. Strain  and Y. Su  and M. Szalay  and G. Vidal  and B. Villalonga  and C. Vollgraff-Heidweiller  and T. White  and Z. Yao  and P. Yeh  and J. Yoo  and A. Zalcman  and Y. Zhang  and N. Zhu  and H. Neven  and D. Bacon  and J. Hilton  and E. Lucero  and R. Babbush  and S. Boixo  and A. Megrant  and Y. Chen  and J. Kelly  and V. Smelyanskiy  and D. A. Abanin  and P. Roushan},
  title         = {{Noise-resilient edge modes on a chain of superconducting qubits}},
  journal       = {Science},
  volume        = {378},
  number        = {6621},
  pages         = {785-790},
  year          = {2022},
  doi           = {10.1126/science.abq5769},
  url           = {https://www.science.org/doi/abs/10.1126/science.abq5769}
}

@article{Smith2016,
  title         = {{Many-body localization in a quantum simulator with programmable random disorder}},
  author        = {Smith, J. and Lee, A. and Richerme, P. and Neyenhuis, B. and Hess, P. W. and Hauke, P. and Heyl, M. and Huse, D. A. and Monroe, C.},
  journal       = {Nature Physics},
  volume        = {12},
  issue         = {10},
  pages         = {907--911},
  year          = {2016},
  publisher     = {Nature Publishing Group},
  doi           = {10.1038/nphys3783},
  url           = {https://doi.org/10.1038/nphys3783}
}

@article{Zhang2017,
  author        = {Zhang, J. and Pagano, G. and Hess, P. W. and Kyprianidis, A. and Becker, P. and Kaplan, H. and Gorshkov, A. V. and Gong, Z.-X. and Monroe, C.},
  year          = {2017},
  title         = {{Observation of a many-body dynamical phase transition with a 53-qubit quantum simulator}},
  journal       = {Nature},
  volume        = {551},
  issue         = {7682},
  pages         = {601--604},
  month         = {Nov},
  publisher     = {Nature Publishing Group},
  doi           = {10.1038/nature24654},
  url           = {https://doi.org/10.1038/nature24654}
}

@article{Bernien2017,
  author        = {Bernien, Hannes and Schwartz, Sylvain and Keesling, Alexander and Levine, Harry and Omran, Ahmed and Pichler, Hannes and Choi, Soonwon and Zibrov, Alexander S. and Endres, Manuel and Greiner, Markus and Vuleti{\'c}, Vladan and Lukin, Mikhail D.},
  year          = {2017},
  title         = {{Probing many-body dynamics on a 51-atom quantum simulator}},
  journal       = {Nature},
  volume        = {551},
  issue         = {7682},
  pages         = {579--584},
  month         = {Nov},
  publisher     = {Nature Publishing Group},
  doi           = {10.1038/nature24622},
  url           = {https://doi.org/10.1038/nature24622}
}

@article{Bluvstein2021,
  author        = {D. Bluvstein  and A. Omran  and H. Levine  and A. Keesling  and G. Semeghini  and S. Ebadi  and T. T. Wang  and A. A. Michailidis  and N. Maskara  and W. W. Ho  and S. Choi  and M. Serbyn  and M. Greiner  and V. Vuleti{\'c}  and M. D. Lukin},
  title         = {{Controlling quantum many-body dynamics in driven Rydberg atom arrays}},
  journal       = {Science},
  volume        = {371},
  number        = {6536},
  pages         = {1355-1359},
  year          = {2021},
  doi           = {10.1126/science.abg2530},
  url           = {https://www.science.org/doi/abs/10.1126/science.abg2530}
}

@article{Ebadi2021,
  title         = {{Quantum phases of matter on a 256-atom programmable quantum simulator}},
  author        = {Ebadi, Sepehr and Wang, Tout T. and Levine, Harry and Keesling, Alexander and Semeghini, Giulia and Omran, Ahmed and Bluvstein, Dolev and Samajdar, Rhine and Pichler, Hannes and Ho, Wen Wei and Choi, Soonwon and Sachdev, Subir and Greiner, Markus and Vuletić, Vladan and Lukin, Mikhail D.},
  journal       = {Nature},
  volume        = {595},
  issue         = {7866},
  pages         = {227--232},
  year          = {2021},
  publisher     = {Nature Publishing Group},
  doi           = {10.1038/s41586-021-03582-4},
  url           = {https://doi.org/10.1038/s41586-021-03582-4}
}

@article{Farrell2024,
  title         = {{Scalable Circuits for Preparing Ground States on Digital Quantum Computers: The Schwinger Model Vacuum on 100 Qubits}},
  author        = {Farrell, Roland C. and Illa, Marc and Ciavarella, Anthony N. and Savage, Martin J.},
  journal       = {PRX Quantum},
  volume        = {5},
  issue         = {2},
  pages         = {020315},
  numpages      = {32},
  year          = {2024},
  month         = {Apr},
  publisher     = {American Physical Society},
  doi           = {10.1103/PRXQuantum.5.020315},
  url           = {https://link.aps.org/doi/10.1103/PRXQuantum.5.020315}
}

@article{Griffin2012,
  title         = {{Scaling Behavior and Beyond Equilibrium in the Hexagonal Manganites}},
  author        = {Griffin, S. M. and Lilienblum, M. and Delaney, K. T. and Kumagai, Y. and Fiebig, M. and Spaldin, N. A.},
  journal       = {Phys. Rev. X},
  volume        = {2},
  issue         = {4},
  pages         = {041022},
  numpages      = {10},
  year          = {2012},
  month         = {Dec},
  publisher     = {American Physical Society},
  doi           = {10.1103/PhysRevX.2.041022},
  url           = {https://link.aps.org/doi/10.1103/PhysRevX.2.041022}
}

@article{Iwamura2024,
  title         = {{Analytical derivation and extension of the anti-Kibble-Zurek scaling in the transverse field Ising model}},
  author        = {Iwamura, Kaito and Suzuki, Takayuki},
  journal       = {Phys. Rev. B},
  volume        = {110},
  issue         = {14},
  pages         = {144102},
  numpages      = {16},
  year          = {2024},
  month         = {Oct},
  publisher     = {American Physical Society},
  doi           = {10.1103/PhysRevB.110.144102},
  url           = {https://link.aps.org/doi/10.1103/PhysRevB.110.144102}
}

@article{Puebla2020,
  title         = {{Universal Anti-Kibble-Zurek Scaling in Fully Connected Systems}},
  author        = {Puebla, Ricardo and Smirne, Andrea and Huelga, Susana F. and Plenio, Martin B.},
  journal       = {Phys. Rev. Lett.},
  volume        = {124},
  issue         = {23},
  pages         = {230602},
  numpages      = {7},
  year          = {2020},
  month         = {Jun},
  publisher     = {American Physical Society},
  doi           = {10.1103/PhysRevLett.124.230602},
  url           = {https://link.aps.org/doi/10.1103/PhysRevLett.124.230602}
}

@article{Singh2023,
  title         = {{Driven one-dimensional noisy Kitaev chain}},
  author        = {Singh, Manvendra and Dhara, Santanu and Gangadharaiah, Suhas},
  journal       = {Phys. Rev. B},
  volume        = {107},
  issue         = {1},
  pages         = {014303},
  numpages      = {11},
  year          = {2023},
  month         = {Jan},
  publisher     = {American Physical Society},
  doi           = {10.1103/PhysRevB.107.014303},
  url           = {https://link.aps.org/doi/10.1103/PhysRevB.107.014303}
}

@misc{Vezvaee2024,
  title         = {{Quantum simulation of Fermi-Hubbard model based on transmon qudit interaction}},
  author        = {Arian Vezvaee and Nathan Earnest-Noble and Khadijeh Najafi},
  year          = {2024},
  eprint        = {2402.01243},
  archivePrefix = {arXiv},
  primaryClass  = {quant-ph},
  url           = {https://arxiv.org/abs/2402.01243}
}

@misc{Gyawali2025,
  title         = {{Observation of disorder-free localization using a (2+1)D lattice gauge theory on a quantum processor}},
  author        = {Gaurav Gyawali and Shashwat Kumar and Yuri D. Lensky and Eliott Rosenberg and Aaron Szasz and Tyler Cochran and Renyi Chen and Amir H. Karamlou and Kostyantyn Kechedzhi and Julia Berndtsson and Tom Westerhout and Abraham Asfaw and Dmitry Abanin and Rajeev Acharya and Laleh Aghababaie Beni and Trond I. Andersen and Markus Ansmann and Frank Arute and Kunal Arya and Nikita Astrakhantsev and Juan Atalaya and Ryan Babbush and Brian Ballard and Joseph C. Bardin and Andreas Bengtsson and Alexander Bilmes and Gina Bortoli and Alexandre Bourassa and Jenna Bovaird and Leon Brill and Michael Broughton and David A. Browne and Brett Buchea and Bob B. Buckley and David A. Buell and Tim Burger and Brian Burkett and Nicholas Bushnell and Anthony Cabrera and Juan Campero and Hung-Shen Chang and Zijun Chen and Ben Chiaro and Jahan Claes and Agnetta Y. Cleland and Josh Cogan and Roberto Collins and Paul Conner and William Courtney and Alexander L. Crook and Sayan Das and Dripto M. Debroy and Laura DeLorenzo and Alexander Del Toro Barba and Sean Demura and Agustin DiPaolo and Paul Donohoe and Ilya Drozdov and Andrew Dunsworth and Clint Earle and Alec Eickbusch and Aviv Moshe Elbag and Mahmoud Elzouka and Catherine Erickson and Lara Faoro and Reza Fatemi and Vinicius S. Ferreira and Leslie Flores Burgos and Ebrahim Forati and Austin G. Fowler and Brooks Foxen and Suhas Ganjam and Robert Gasca and William Giang and Craig Gidney and Dar Gilboa and Raja Gosula and Alejandro Grajales Dau and Dietrich Graumann and Alex Greene and Jonathan A. Gross and Steve Habegger and Michael C. Hamilton and Monica Hansen and Matthew P. Harrigan and Sean D. Harrington and Stephen Heslin and Paula Heu and Gordon Hill and Jeremy Hilton and Markus R. Hoffmann and Hsin-Yuan Huang and Ashley Huff and William J. Huggins and Lev B. Ioffe and Sergei V. Isakov and Evan Jeffrey and Zhang Jiang and Cody Jones and Stephen Jordan and Chaitali Joshi and Pavol Juhas and Dvir Kafri and Hui Kang and Trupti Khaire and Tanuj Khattar and Mostafa Khezri and Mária Kieferová and Seon Kim and Paul V. Klimov and Andrey R. Klots and Bryce Kobrin and Alexander N. Korotkov and Fedor Kostritsa and John Mark Kreikebaum and Vladislav D. Kurilovich and David Landhuis and Tiano Lange-Dei and Brandon W. Langley and Pavel Laptev and Kim-Ming Lau and Loick LeGuevel and Justin Ledford and Joonho Lee and Kenny Lee and Brian J. Lester and Wing Yan Li and Alexander T. Lill and Wayne Liu and William P. Livingston and Aditya Locharla and Daniel Lundahl and Aaron Lunt and Sid Madhuk and Ashley Maloney and Salvatore Mandrà and Leigh S. Martin and Steven Martin and Orion Martin and Cameron Maxfield and Jarrod R. McClean and Matt McEwen and Seneca Meeks and Anthony Megrant and Xiao Mi and Kevin C. Miao and Amanda Mieszala and Sebastian Molina and Shirin Montazeri and Alexis Morvan and Ramis Movassagh and Charles Neill and Ani Nersisyan and Michael Newman and Anthony Nguyen and Murray Nguyen and Chia-Hung Ni and Murphy Yuezhen Niu and William D. Oliver and Kristoffer Ottosson and Alex Pizzuto and Rebecca Potter and Orion Pritchard and Leonid P. Pryadko and Chris Quintana and Matthew J. Reagor and David M. Rhodes and Gabrielle Roberts and Charles Rocque and Nicholas C. Rubin and Negar Saei and Kannan Sankaragomathi and Kevin J. Satzinger and Henry F. Schurkus and Christopher Schuster and Michael J. Shearn and Aaron Shorter and Noah Shutty and Vladimir Shvarts and Volodymyr Sivak and Jindra Skruzny and Spencer Small and W. Clarke Smith and Sofia Springer and George Sterling and Jordan Suchard and Marco Szalay and Alex Sztein and Douglas Thor and M. Mert Torunbalci and Abeer Vaishnav and Sergey Vdovichev and Guifre Vidal and Catherine Vollgraff Heidweiller and Steven Waltman and Shannon X. Wang and Theodore White and Kristi Wong and Bryan W. K. Woo and Cheng Xing and Z. Jamie Yao and Ping Yeh and Bicheng Ying and Juhwan Yoo and Noureldin Yosri and Grayson Young and Adam Zalcman and Yaxing Zhang and Ningfeng Zhu and Nicholas Zobrist and Sergio Boixo and Julian Kelly and Erik Lucero and Yu Chen and Vadim Smelyanskiy and Hartmut Neven and Dmitry Kovrizhin and Johannes Knolle and Jad C. Halimeh and Igor Aleiner and Roderich Moessner and Pedram Roushan},
  year          = {2025},
  eprint        = {2410.06557},
  archivePrefix = {arXiv},
  primaryClass  = {quant-ph},
  url           = {https://arxiv.org/abs/2410.06557}
}

@misc{Hayata2025,
  title         = {{Digital quantum simulation of many-body localization crossover in a disordered kicked Ising model}},
  author        = {Tomoya Hayata and Kazuhiro Seki and Seiji Yunoki},
  year          = {2025},
  eprint        = {2510.01983},
  archivePrefix = {arXiv},
  primaryClass  = {quant-ph},
  url           = {https://arxiv.org/abs/2510.01983}
}
\let\addcontentsline\oldaddcontentsline

\clearpage

\setcounter{page}{1}
\setcounter{section}{0}
\setcounter{equation}{0}
\setcounter{figure}{0}

\renewcommand{\thepage}{S\arabic{page}}
\renewcommand{\theequation}{S.\arabic{equation}}
\renewcommand{\theHequation}{S.\arabic{equation}}
\renewcommand{\thefigure}{S\arabic{figure}}
\renewcommand{\theHfigure}{S\arabic{figure}}

\onecolumngrid

\begin{center}
{\large \textbf{Supplementary Materials for \\ ``\ourtitle"}}\\
\vspace{0.5cm}
Brendan Rhyno$^{1,2}$,
Swarnadeep Majumder$^{3,4}$, 
Smitha Vishveshwara$^{2}$,
Khadijeh Najafi$^{4,3}$\\ 
\vspace{0.25cm}
\textit{$^1$\Hannover}\\
\textit{$^2$\UIUC} \\
\textit{$^3$\ibmyorktown}\\
\textit{$^4$\ibmcambridge}\\
\end{center}

\tableofcontents

\vspace{1.0cm}

\twocolumngrid

\section{Quantum Kibble-Zurek mechanism under continuous quantum nondemolition measurements of the instantaneous Hamiltonian}
\label{SI:QND}

In this section, we theoretically probe how the QKZ mechanism is modified under continuous quantum nondemolition measurements of the instantaneous Hamiltonian.
Although not of direct relevance to the IBM Quantum hardware, this noise channel offers a concrete example of how universal dynamics can be reshaped by decoherence.
First, we review how quenches across a quantum critical point yield distinct universal scaling relations depending on the strength of nondemolishing noise.
We then consider the consequences for the one-dimensional transverse-field Ising model, where the many-body dynamics can be approximately mapped onto a collection of two-state Landau-Zener-like problems, and we formally compute relevant equal-time spin correlation functions.
We conclude by outlining our numerical method to compute these correlation functions.

\subsection{Quenching across a quantum critical point}

Under the Born and Markov approximations, the density operator of an open quantum system evolves according to a Gorini-Kossakowski-Sudarshan-Lindblad equation~\cite{Nielsen2001,Miessen2023}.
Here, we consider a specific type of noise channel that was recently shown to modify critical scaling relations.
In particular, we investigate the case in which the environment continuously monitors the instantaneous energy eigenstates of the Hamiltonian.
These continuous quantum nondemolition (QND) measurements of the instantaneous Hamiltonian lead to the following master equation ($\hbar = 1$):
\begin{equation}
    \label{eq:QND_Linblad_master_equation}
    \partial_t \RR(t) = -i [\HH(t), \RR(t)] - \lambda [\HH(t),[\HH(t) , \RR (t) ] ]
    ,
\end{equation}
where $\lambda \ge 0$ represents the QND measurement strength~\cite{Kuo2021}.
This is an example of the general master equation with jump operator $\LL(t) = \sqrt{2 \lambda} \HH(t)$.
Unitary dynamics corresponds to the case that $\lambda=0$.
In contrast, the system has a tendency to decohere into its energy eigenbasis for large $\lambda$.

We imagine the standard quantum Kibble-Zurek (QKZ) scenario, in which the system is initialized in the ground state of the Hamiltonian. Then a parameter $\epsilon$ in the Hamiltonian is dynamically tuned towards a quantum critical point (QCP) that occurs at $\epsilon=0$.
In the vicinity of the QCP, the control parameter can be expanded as $\epsilon(t) \propto t/\tau_Q$ as $t \to 0$, where $\tau_Q$ is a characteristic time scale called the \textit{quench time}.
For a system evolving according to Eq.~(\ref{eq:QND_Linblad_master_equation}), it was shown in~\cite{Kuo2021} that a Kibble-Zurek-type scaling argument can be made.
Depending on the strength of QND measurement noise, $\lambda$, one can identify different freeze-out time scales at which the system falls out of equilibrium.
For weak decoherence, $\lambda \to 0$, the term responsible for unitary evolution in the master equation dominates, and standard QKZ scaling is recovered.
However, for strong decoherence, $\lambda \to \infty$, the continuous QND measurement term dominates and the freeze-out time $\bar{t}$ is set by the time at which the tuning rate, $\dot{\epsilon}(t) / \epsilon(t) \sim 1/t$ as $t \to 0$,
is related to the many-body spectral gap, $\Delta(t) \propto |\epsilon(t)|^{z\nu} \propto |t/\tau_Q|^{z\nu}$ as $t \to 0$, via~\cite{Kuo2021}
\begin{equation}
    \left| \frac{\dot{\epsilon}(\bar{t})}{\epsilon(\bar{t})} \right| \equiv
    \lambda \Delta(t)^2
    \to \bar{t} \propto \tau_Q^{2z\nu/(1+2z\nu)}
    ,
\end{equation}
with the corresponding correlation length,
\begin{equation}
    \bar{\xi} \equiv \xi(\bar{t}) \to \bar{\xi} \propto \tau_Q^{\nu/(1+2z\nu)}
    .
\end{equation}
This decoherent scaling still involves the quench time and universal critical exponents. However, the algebraic combination of critical exponents has been modified from the closed system case.
For large $\lambda$, one can then make a modified Kibble-Zurek scaling hypothesis for correlation functions of local operators:
\begin{equation}
    \label{eq:QND_QKZ_scaling_hypothesis}
    \langle  O(x) \rangle_t
    = \bar{\xi}^{-\Delta_{O}} \mathcal{F}_O(t/\bar{t},x/\bar{\xi})
    ,
\end{equation}
where $\Delta_O$ is the scaling dimension of the operator~\cite{Francuz2016}. Inserting the freeze-out time and correlation length above, we obtain
\begin{equation}
    \langle  O(x) \rangle_t
    = \tau_Q^{-\Delta_{O} \nu/(1+2z\nu)}
    \mathcal{G}_{O} \left( \frac{ t }{ \tau_Q^{2z\nu/(1+2z\nu)} } , \frac { x }{\tau_Q^{\nu/(1+2z\nu)} } \right)
    .
\end{equation}
Thus, as $\lambda\to\infty$, there is a predicted data collapse at the QCP for $\langle  O(x) \rangle_{t=0} \tau_Q^{\Delta_{O} \nu/(1+2z\nu)}$ versus $x/\tau_Q^{\nu/(1+2z\nu)}$.
For a quantum spin--1/2 system, this decoherent QKZ mechanism predicts that the equal-time connected ZZ correlation function,
$C^{zz}_{ij}(t) = \langle \sig^z_i \sig^z_j \rangle_t - \langle \sig^z_i \rangle_t \langle \sig^z_j \rangle_t$
, will exhibit a data collapse for $C^{zz}_{ij}(t=0) \tau_Q^{b_\text{QND}}$ versus $|i-j|/\tau_Q^{a_\text{QND}}$, where
\begin{equation}
    \label{eq:QND_QKZ_scaling}
    a_\mathrm{QND} \equiv \nu / (1 + 2 z \nu)
    \quad,\quad
    b_\mathrm{QND} \equiv \Delta_{zz} \nu / (1 + 2 z \nu)
    .
\end{equation}
[Note: due to translation invariance in the theory model, the site-averaged correlation function discussed in the main text, Eq.~(\ref{eq:site_averaged_connected_2pt_fn}), is simply $C(t,x) = C^{zz}_{i,i+x}(t)$.]
For the one-dimensional Ising universality class, these exponents take on the values  $a_\text{QND} = 1/3 = 0.333\dots$ and $b_\text{QND} = 1/12 = 0.083\dots$, which can be compared to the $\lambda=0$ exponents, $a_\text{QKZ} = 1/2 = 0.5$ and $b_\text{QKZ} = 1/8 = 0.125$.

\subsection{Decoherent quenches in the one-dimensional transverse-field Ising model}
\label{SI:QND_TFI}

Consider the periodic TFIM Eq.~(\ref{eq:TFIM}) with $N$ sites, which we take to be even for convenience.
The system undergoes the quench protocol described in the main text, but here we consider decoherence brought about by continuous QND measurements of the instantaneous Hamiltonian.
The initial paramagnetic ground state density operator evolves in time according to the master equation Eq.~(\ref{eq:QND_Linblad_master_equation}).

It is helpful to map the problem onto spinless fermions using a standard Jordan-Wigner transformation~\cite{Sachdev2001}.
Namely, we express the spin--1/2 operators in terms of fermionic creation and annihilation operators $\cd_i$ and $\cc_i$ respectively, using
\begin{subequations}
\begin{align}
    \label{eq:JW_transformation_TFI}
    \sig^x_i &= 1 - 2 \cd_i \cc_i
    , \\
    \sig^y_i &= i(\cd_i-\cc_i) \prod_{j<i} (1 - 2 \cd_j \cc_j )
    , \\
    \sig^z_i &= -(\cd_i+\cc_i) \prod_{j<i} (1 - 2 \cd_j \cc_j )
    .
\end{align}
\end{subequations}
With these definitions, the vacuum state of the Jordan-Wigner fermions $| 0 \rangle_c$ is the state with all spins pointed up along the $x$-axis: $| 0 \rangle_c = | + \rangle^{\otimes N}$. The TFIM Hamiltonian then becomes
\begin{align}
    \HH(t) &=
    -N h(t)
    \nonumber\\
    &
    + 2 h(t) \sum_{i=1}^{N} \cd_i \cc_i
    - J(t) \sum_{i=1}^{N-1} (\cd_i \cc_{i+1} + \cd_i \cd_{i+1} + \text{h.c})
    \nonumber\\
    &+ J(t) \PP (\cd_N \cc_1 + \cd_N \cd_1 + \text{h.c})
    \label{eq:TFI_JW}
    .
\end{align}
The last term appears when using PBCs and involves the parity operator $\PP \equiv \prod_{i} (1 - 2 \cd_i \cc_i)$, which has the property $\PP \, \cc_i \, \PP = - \cc_i$. Because the Hamiltonian is parity symmetric, $[\HH(t),\PP]=0$, one can write~\cite{Dziarmaga2010,Francuz2016}
\begin{equation}
    \HH(t) = \PP^+ \HH^+(t) \PP^+ +  \PP^- \HH^-(t) \PP^-
    ,
\end{equation}
where $\PP^{\pm} \equiv \frac{1}{2}(I\pm\PP)$ is the projector onto the even/odd parity sectors of Hilbert space and the Hamiltonian within each sector is given by
\begin{align}
    \HH^{\pm}(t) &=
    -N h(t)
    \nonumber\\
    &
    + 2 h(t) \sum_{i=1}^{N} \cd_i \cc_i
    - J(t) \sum_{i=1}^{N} (\cd_i \cc_{i+1} + \cd_i \cd_{i+1} + \text{h.c})
    ,
\end{align}
where $\cc_{N+1} \equiv \mp \cc_1$.
In particular, in the even-parity sector the fermions can be treated as having antiperiodic boundary conditions (ABCs) and in the odd-parity sector they can be treated as having PBCs. 

In the fermionic description, the initial density operator is the vacuum projector, which has even parity. Time-evolving the system with Eq.~(\ref{eq:QND_Linblad_master_equation}) preserves this parity, hence we can replace the full Hamiltonian in the master equation with the operator in the even parity sector. Namely, we treat $\HH^+$ as the generator of time evolution~\cite{Francuz2016}, and the master equation for the density operator becomes:
\begin{equation}
    \partial_t \RR(t) = -i [\HH^+(t), \RR(t)] - \lambda [\HH^+(t),[\HH^+(t) , \RR(t)]]
    .
    \label{eq:Master_eq_even_parity}
\end{equation}

\subsubsection{Instantaneous ground state}
Before moving further, it is instructive to study the adiabatic ground state of the system. The Hamiltonian can be diagonalized via a standard Bogoliubov-de Gennes (BdG) transformation.
Diagonalization proceeds most efficiently in momentum space
\begin{equation}
    \cc_k \equiv \frac{1}{\sqrt{N}} \sum_{j=1}^{N} e^{-i kj} \cc_j
    \quad\leftrightarrow\quad
    \cc_j = \frac{1}{\sqrt{N}} \sum_k e^{+i kj} \cc_k
    ,
    \label{eq:c_k}
\end{equation}
where, appropriate for ABCs, we define $k = \frac{2\pi}{N}(n+1/2)$, with $n\in\mathbb{Z}$  and wavevectors remaining in the first Brillouin zone: $k\in(-\pi,\pi)$.
In momentum space, the Hamiltonian becomes
\begin{align}
    \HH^{+}(t)
    &=
    -N h(t)
    + \sum_{0<k<\pi} \Big( 2 h(t) - 2 J(t) \cos(k) \Big)
    \nonumber\\
    &
    + \sum_{0<k<\pi} \Psi^\dag_k H^{\text{BdG}}_k(t) \Psi_k
    \label{eq:TFI_Nambu}
    ,
\end{align}
where we have used that $N$ is even and have defined the Nambu spinor and BdG Hamiltonian matrix as:
\begin{equation}
    \Psi_k \equiv
    \begin{pmatrix}
    \cc_k \\ \cd_{-k} \end{pmatrix}
    \quad , \quad
    H^\text{BdG}_k(t) \equiv
    \vec{h}_k(t) \cdot \vec{\tau}
    ,
\end{equation}
with $\tau^\alpha$ being Pauli matrices in Nambu space and $\vec{h}_k(t)$ being a pseudo-magnetic field,
\begin{subequations}
\begin{align}
    h^x_k(t) &\equiv 0
    , \\
    h^y_k(t) &\equiv 2J(t) \sin(k)
    , \\
    h^z_k(t) &\equiv 2h(t) - 2J(t) \cos(k)
    .
\end{align}
\end{subequations}

The instantaneous Hamiltonian in diagonal form is
\begin{align}
    \HH^{+}(t) &=
    -N h(t)
    - \sum_{0<k<\pi} \Big( |\vec{h}_k(t)| - h^z_k(t) \Big)
    \nonumber\\
    &
    + \sum_{0<k<\pi} |\vec{h}_k(t)| \Big( \gam^\dag_k(t) \gam_k(t) + \gam^\dag_{-k}(t) \gam_{-k}(t) \Big)
    \label{eq:TFI_diagonal}
    ,
\end{align}
where fermionic quasiparticle creation operators $\gam^\dag_k(t)$ have been introduced defined via a BdG transformation
\begin{align}
    \gam^\dag_k(t) &\equiv u_k(t) \cd_k + v_k(t) \cc_{-k}
    ,\\
    u_k(t)
    &\equiv \sqrt{\frac{|\vec{h}_k(t)|+h^z_k(t)}{2|\vec{h}_k(t)|}}
    \label{eq:u_k}
    ,\\
    v_k(t)
    &\equiv e^{i \text{arg}(h^x_k(t)+ih^y_k(t))} \sqrt{\frac{|\vec{h}_k(t)|-h^z_k(t)}{2|\vec{h}_k(t)|}}
    \label{eq:v_k}
    .
\end{align}
The instantaneous vacuum state of the BdG quasiparticles $| \text{vac},t \rangle$ can be constructed from the original fermion vacuum state $| 0 \rangle_c$ as
\begin{equation}
    | \text{vac},t \rangle
    = \prod_{0<k<\pi} (u_k^*(t) - v_k^*(t) \cd_k \cd_{-k}) | 0 \rangle_c
    .
    \label{eq:TFI_vacuum}
\end{equation}

From the BdG quasiparticle vacuum, combined with the fact that the vacuum projector of the Jordan-Wigner fermions can be written as $| 0 \rangle_c {}_c\langle 0 | = \prod_i \cc_i \cd_i = \prod_k \cc_k \cd_k$,
one finds that the instantaneous ground state density operator of $\HH^+(t)$ is given by
\begin{align}
    \RR_\text{GS}(t)
    &= \prod_{0<k<\pi} \big[u_k^*(t) - v_k^*(t) \cd_k \cd_{-k}\big] | 0 \rangle_k {}_k\langle 0 | \big[u_k(t)
    \nonumber\\
    &- v_k(t) \cc_{-k}\cc_k \big]
    ,
\end{align}
where $| 0 \rangle_k {}_k\langle 0 | \equiv \cc_k \cd_k \cc_{-k} \cd_{-k}$ projects onto the subspace of states with unoccupied $k$ and $-k$ modes.

Thus, the instantaneous ground state density operator is factorized into commuting operators for each $(k,-k)$ pair, each of which connects unoccupied states and Cooper-paired states:
\begin{align}
    | 0 \rangle_k {}_k\langle 0 |
    &    ,\\
    | 0 \rangle_k {}_k\langle 1 |
    &\equiv
    | 0 \rangle_k {}_k\langle 0 |
    \cc_{-k} \cc_{k}
    ,\\
    | 1 \rangle_k {}_k\langle 0 |
    &\equiv
    \cd_{k} \cd_{-k}
    | 0 \rangle_k {}_k\langle 0 |
    ,\\
    | 1 \rangle_k {}_k\langle 1 |
    &\equiv
    \cd_{k} \cd_{-k}
    | 0 \rangle_k {}_k\langle 0 |
    \cc_{-k} \cc_{k}
    .
\end{align}
With this notation, the instantaneous ground state density operator then takes the form
\begin{align}
    \RR_\text{GS}(t)
    &= \prod_{0<k<\pi} | n \rangle_k {}_k\langle n' | \left[ \varrho_k^{\text{GS}}(t) \right]_{nn'}
    ,\\
    \varrho_k^{\text{GS}}(t)
    &\equiv
    \begin{pmatrix}
    |u_k(t)|^2 & - u_k^*(t) v_k(t) \\
    - u_k(t) v_k^*(t) & |v_k(t)|^2
    \end{pmatrix}
    ,
\end{align}
where $n,n' \in \{0,1\}$ counts the number of Cooper pairs for each $(k,-k)$ pair and there is an implied summation on repeated $n$ and $n'$.

\subsubsection{Ansatz for the density operator}

Based on the form of the instantaneous ground state density operator, we make an ansatz and try to solve Eq.~(\ref{eq:Master_eq_even_parity}) using an operator which has the following form:
\begin{equation}
    \label{eq:ansatz_rho}
    \RR(\{ \varrho_k(t) \}) \equiv \prod_{0<k<\pi} 
    | n \rangle_k {}_k\langle n' | \left[ \varrho_k(t) \right]_{nn'}
    .
\end{equation}
Here, $\varrho_k(t)$ represents a 2x2 density matrix in the subspace of unoccupied and Cooper-paired states for each $(k,-k)$ pair, and is taken to obey $\varrho_k^\dag(t) = \varrho_k(t)$, $\tr( \varrho_k(t)) = 1$, and $\tr(\varrho_k^2(t)) \le 1$.
Plugging this ansatz into each term of the master equation, Eq.~(\ref{eq:Master_eq_even_parity}), separately gives
\begin{align}
    \partial_t \RR(\{ \varrho_k(t) \})
    &=
    \sum_{0<k<\pi}
    \RR(\{ \partial_t \varrho_k(t) , \varrho_{k' \neq k}(t) \})
    ,\\
    -i [\HH^+(t),\RR(\{ \varrho_k(t) \})]
    &=
    -i 
    \sum_{0<k<\pi}
    \RR(\{ \tilde{\varrho}_k(t) , \varrho_{k' \neq k}(t) \})
    ,\\
    - \lambda [\HH^+(t),[\HH^+(t),\RR&(\{ \varrho_k(t) \})]]
    \nonumber\\
    &=
    - \lambda 
    \sum_{0<k<\pi}
    \bigg[
    \RR(\{ \tilde{\tilde{\varrho}}_k(t) , \varrho_{k' \neq k}(t) \})
    \nonumber\\
    +
    \sum_{0<k' \neq k<\pi}&
    \RR(\{ \tilde{\varrho}_k(t) , \tilde{\varrho}_{k'}(t), \varrho_{k'' \neq k,k'}(t) \})
    \bigg]
    ,
    \label{eq:QND_ansatz_double_commutator_term}
\end{align}
where we define the following operation on 2x2 density matrices:
\begin{equation}
    \tilde{\varrho}_k(t) \equiv [- H^\text{BdG}_k(t), \varrho_k(t)]
    .
\end{equation}

Owing to the linearity of the ansatz,
$
    \RR(\{a A_k + b B_k , \varrho_{k' \neq k} \})
    = a \, \RR(\{A_k, \varrho_{k' \neq k}\}) + b \, \RR(\{B_k, \varrho_{k' \neq k}\})
    ,
$
$a,b\in\mathbb{C}$,
one can see that the ansatz \textit{almost} solves Eq.~(\ref{eq:Master_eq_even_parity}), provided the 2x2 density matrices evolve according to
\begin{equation}
    \partial_t \varrho_k(t) = -i \, \tilde{\varrho}_k(t) - \lambda \, \tilde{\tilde{\varrho}}_k(t)
    .
    \label{eq:Master_eq_2x2}
\end{equation}
However, for $\lambda \neq 0$, the last term in Eq.~(\ref{eq:QND_ansatz_double_commutator_term}) couples different positive $k$ and $k'$ modes and prevents a true solution.
Here, we follow similar works and \textit{neglect} the $k$, $k'$ mode mixing in Eq.~(\ref{eq:QND_ansatz_double_commutator_term})~\cite{Nalbach2015,Dutta2016,Kuo2021}.
See~\cite{Nalbach2015,Dutta2016} for a discussion on the validity of this assumption.
Under this approximation, we consider a density operator of the form Eq.~(\ref{eq:ansatz_rho}) whose 2x2 density matrices in the subspace of unoccupied and Cooper-paired states for each $(k,-k)$ pair evolves according to Eq.~(\ref{eq:Master_eq_2x2}).

\subsubsection{Parameterization on the Bloch ball}

Any 2x2 density matrix can be parameterized in terms of a vector residing inside or on the surface of a unit 2-sphere:
\begin{equation}
    \label{eq:Bloch_vector}
    \varrho_k(t) \equiv \frac{1}{2} (1 + \vec{n}_k(t) \cdot \vec{\tau} ),
\end{equation}
where $n^\alpha_k(t) \in \mathbb{R}$ is a component of the Bloch vector with $|\vec{n}_k(t)| \le 1$~\cite{Nielsen2001}.
In terms of the Bloch vector, the master equation for the density matrix, Eq.~(\ref{eq:Master_eq_2x2}), becomes~\cite{Kuo2021}
\begin{align}
    \partial_t \vec{n}_k(t)
    =
    &-2 \, \vec{h}_k(t) \times \vec{n}_k(t)
    \nonumber\\
    &+4 \lambda \, \vec{h}_k(t) \times (\vec{h}_k(t) \times \vec{n}_k(t))
    \label{eq:Master_eq_Bloch_vector}
    .
\end{align}
[Note: in the main text, we define $\vec{B}_k(t) = -2 \, \vec{h}_k(t)$ to condense the above expression.]

The Bloch vector gives a particularly nice geometric understanding of the dynamics of the system during a decoherent quench. For instance, one finds that the residual energy (energy above the instantaneous ground state) is given by:
\begin{equation}
    E_\text{res}(t) = \sum_{0<k<\pi} \Big( |\vec{h}_k(t)| -\vec{h}_k(t) \cdot \vec{n}_k(t) \Big)
    .
\end{equation}
Thus, the evolution of the system is adiabatic only if the Bloch vector traces out the path given by $\vec{n}_k(t) = \vec{h}_k(t) / |\vec{h}_k(t)|$.

\subsubsection{Equal-time fermionic correlators}

One can compute various fermionic correlators using the density operator in Eq.~(\ref{eq:ansatz_rho}). For instance, one obtains the following equal-time 2-point fermionic correlation functions ($k>0$):
\begin{align}
    \langle\cc_{-k} \cd_{-k}\rangle_t &= [\varrho_k(t)]_{00}
    = \frac{1}{2} \left( 1 + n^z_k(t) \right)
    ,\\
    \langle\cd_{k} \cd_{-k}\rangle_t &= [\varrho_k(t)]_{01}
    = \frac{1}{2} \left( n^x_k(t) - i n^y_k(t) \right)
    ,\\
    \langle\cc_{-k} \cc_{k}\rangle_t &= [\varrho_k(t)]_{10}
    = \frac{1}{2} \left( n^x_k(t) + i n^y_k(t) \right)
    ,\\
    \langle\cd_{k} \cc_{k}\rangle_t &= [\varrho_k(t)]_{11}
    = \frac{1}{2} \left( 1 - n^z_k(t) \right)
    .
\end{align}
The Bloch vector is then related to the 2-point fermionic correlators via $\vec{n}_k(t) = \langle \Psi_k^\dagger \, \tau^{x} \vec{\tau} \tau^{x}  \,\Psi_k \rangle_t$.
From these expressions, one can determine various physically relevant quantities such as the number of quasiparticle excitations and energy.
Fourier transforming also gives one access to spatially-resolved correlation functions:
\begin{subequations}
\label{eq:JW_2pt_fns}
\begin{align}
    \langle\cc_j \cd_l\rangle_t &=
    \frac{2}{N} \sum_{0<k<\pi} \cos(k(j-l)) [\varrho_k(t)]_{00}
    ,\\
    \langle\cd_j \cd_l\rangle_t &=
    -\frac{2i}{N} \sum_{0<k<\pi} \sin(k(j-l)) [\varrho_k(t)]_{01}
    ,\\
    \langle\cc_j \cc_l\rangle_t &=
    -\frac{2i}{N} \sum_{0<k<\pi} \sin(k(j-l)) [\varrho_k(t)]_{10}
    ,\\
    \langle\cd_j \cc_l\rangle_t &=
    \frac{2}{N} \sum_{0<k<\pi} \cos(k(j-l)) [\varrho_k(t)]_{11}
    .
\end{align}
\end{subequations}

\subsubsection{Equal-time spin correlation functions}

Our interests here lie in computing equal-time spin correlation functions. These can be constructed using the fermionic correlators above, Eq.~(\ref{eq:JW_2pt_fns}). Using the Jordan-Wigner transformation, Eq.~(\ref{eq:JW_transformation_TFI}), the local magnetization along the $x$-axis is
\begin{equation}
    M^x_i(t) = \langle\sig^x_i\rangle_t = 1 - 2\langle\cd_i \cc_i\rangle_t
    ,
\end{equation}
with $M^y_i(t) = M^z_i(t) = 0$ as these expectation values only involve expressions with an odd number of fermions.

Moving onto equal-time connected spin-spin correlation functions, we use Wick's theorem to write the correlators as products over fermionic 2-point functions~\cite{Sachdev2001}.
For the equal-time connected XX correlation function one obtains
\begin{align}
    C^{xx}_{ij}(t) &= \langle\sig^x_i\sig^x_j\rangle_t - \langle\sig^x_i\rangle_t \langle\sig^x_j\rangle_t
    \nonumber\\
    &=
    4 \left[
    \langle\cd_i \cc_j\rangle_t \langle\cc_i \cd_j\rangle_t
    -
    \langle\cd_i \cd_j\rangle_t \langle\cc_i \cc_j\rangle_t
    \right]
    .
\end{align}
For the equal-time connected YY and ZZ correlation functions, one needs to be more careful due to the presence of the Jordan-Wigner string. Focusing on the ZZ correlator, and following for instance~\cite{Sachdev2001}, one can write
\begin{align}
    &C^{zz}_{ij}(t) = \langle\sig^z_i\sig^z_j\rangle_t - \cancel{\langle\sig^z_i\rangle_t} \cancel{\langle\sig^z_j\rangle_t}
    \\
    &=
    \langle (\cd_i + \cc_i) \prod_{m<i}(1-2\cd_m \cc_m) \prod_{n<j} (1-2\cd_n\cc_n) (\cd_j + \cc_j) \rangle {}_t
    ,
\end{align}
where we used that expectation values of parity-switching operators vanish.
Assuming $j>i$ for concreteness and writing $1-2\cd_i \cc_i = (\cd_i + \cc_i)(\cd_i-\cc_i)$, this becomes
\begin{equation}
    C^{zz}_{ij}(t)
    =
    \langle (\cd_i - \cc_i) \prod_{m={i+1}}^{j-1}(\cd_m + \cc_m) (\cd_m - \cc_m) (\cd_j + \cc_j) \rangle {}_t
    .
\end{equation}
Applying Wick's theorem, this reduces to a Pfaffian whose elements are combinations of 2-point fermionic correlators~\cite{Lieb1961,Barouch1971,McCoy1971,Francuz2016}.

\subsection{Numerical procedure for computing equal-time spin correlation functions}
\label{SI:theory_numerics}

We numerically solve the master equation, Eq.~(\ref{eq:Master_eq_Bloch_vector}), to obtain the components of the Bloch vector $\vec{n}_k(t)$ for each $k$ mode.
In particular, we utilize \texttt{SciPy}~\cite{SciPy2020} to integrate the system of ordinary differential equations with initial data $\vec{n}_k(0)=(0,0,1)$.
We then Fourier transform the Bloch vector components to find the equal-time fermionic 2-point correlation functions in Eq.~(\ref{eq:JW_2pt_fns}).
Finally, we compute Pfaffians of matrices constructed from these fermionic correlators to determine the equal-time connected ZZ correlation function. 
The Pfaffians were computed in python using \texttt{pfapack}~\cite{Wimmer2012}.

\section{Extracting scaling exponents at the quantum critical point}
\label{SI:fitting}

The (standard) QKZ mechanism predicts a data collapse at the QCP for $C(0,x) \tau_Q^{b}$ versus $x / \tau_Q^{a}$ using exponents discussed in the main text.
One also finds such scaling relations exist in the case of strong continuous quantum nondemolition measurements where the exponents are modified, but still contain information on the Ising universality class (see Sec.~\ref{SI:QND}).
In studying quenches to the QCP using IBM Quantum hardware, we are interested in seeing if such scaling behavior emerges.
Thus, given a set of correlation function data $C(0,x)$ obtained from (numerical or hardware) quenches to the QCP, each characterized by a quench time $\tau_Q$, we employ a robust optimization scheme similar to the one developed in~\cite{Dupont2022} to search for exponents $(a,b)$ that lead to the best data collapse in $C(0,x) \tau_Q^{b}$ versus $x / \tau_Q^{a}$.
In particular, following~\cite{Dupont2022}, we sweep the values of $a$ and $b$ over a square-grid and, for each point $(a,b)$, we fit the rescaled data to an exponentially decaying function of the form
\begin{equation}
    f(y) = e^{- p_{-1} y} \sum_{m=0}^{M} p_m y^m
    ,
\end{equation}
with fit parameters $p_{-1},p_0,\dots,p_M$.
We used $M=4$, which leads to 6 fit parameters for a given correlation function data set and each $(a,b)$ pair.
We then take $(a_\mathrm{fit},b_\mathrm{fit})$ to be the $(a,b)$ pair whose fit yields the smallest root mean square error (RMSE).
As our primary interest lies in the flow of scaling exponents away from standard QKZ values as the system is subjected to a combination of noise, finite-size effects, and Trotterization, we consider a coarse square lattice of $(a,b)$ points with lattice spacing $0.025$.
Hence, all reported best-fit scaling exponents are only resolvable up to $\pm 0.025$.

\section{Digital quantum simulation details}
\label{SI:quantum_simulation}

In this section, we explore our experimental procedures and methods.
We begin with a detailed explanation of the quantum circuits used to simulate Trotterized quenches in the one-dimensional transverse-field Ising model.
We then discuss our various choices for simulating Kibble-Zurek dynamics on the IBM Quantum hardware, such as the choice of Trotter step, sample shot (along with a discussion on error analysis) and error mitigation.
We conclude by showing additional experimental data from our digital quantum simulations.

\subsection{Unitary time evolution in digital quantum simulators}
\label{SI:trotter_step}

For a closed quantum system, the unitary time evolution operator, denoted as $\UU(t_f,t_i)$, translates the system from initial time $t_i$ to final time $t_f$. The operator can be expressed using the time-ordered exponential ($\hbar = 1$):
\begin{equation}
    \label{eq:TOE}
    \UU(t_f,t_i)
    = \mathcal{T} \exp\left( -i \int_{t_i}^{t_f} dt \HH(t) \right)
    .
\end{equation}
One can approximate this expression with a finite number of time steps using~\cite{Poulin2011,Smith2019,Dupont2022,Miessen2024}:
\begin{equation}
    \label{eq:discrete_time_evolution}
    \UU(t_f,t_i)
    \sim
    e^{-i dt \HH(t_{P})}
    \cdots e^{-i dt \HH(t_1)}
    \text{ , as }P\to\infty
    ,
\end{equation}
where $t_s \equiv t_i + s d t$ with $s=1,\dots,P$ and $d t \equiv (t_f - t_i) / P$.
For a finite number of time steps $P$, Eq.~(\ref{eq:discrete_time_evolution}) only approximates Eq.~(\ref{eq:TOE}) as it turns the Hamiltonian parameters into piecewise constant functions of time.
In practice, we fix a value for the time-difference $d t$ and then take time steps $s=1,\dots,P$ forward to discrete time $t_s = t_i + s d t$.

To implement a single time step forward with the operator $\exp(-i dt \HH(t_s))$, we employ a first-order Suzuki-Trotter decomposition~\cite{Trotter1959,Suzuki1993,Lloyd1996,Zalka1998}:
\begin{equation}
    \label{eq:Trotter}
    e^{-i dt ({A} + {B})} = e^{-i dt {A}} e^{-i dt {B}} + \mathcal{O}(d t^2)
    .
\end{equation}
For the TFIM Hamiltonian, Eq.~(\ref{eq:TFIM}), we set ${A}(t) \equiv - h(t) \sum_{i} \sig^x_i$ and ${B}(t) \equiv - J(t) \sum_{i} \sig^z_i \sig^z_{i+1}$~\cite{Dupont2022,Miessen2024}.
The unitary $\exp(-i dt {A})$ simply involves a layer of rotation gates about the $x$-axis.
For a single qubit at site $j$, this rotation gate is denoted by the circuit
\begin{equation}
    \label{eq:X_unitary}
    R^x_j(\theta)
    \equiv e^{-i \frac{\theta}{2} \sig^{x}_j}
    =
    \begin{quantikz}
    \lstick{$j$}  & \gate{R_X(\theta)}   &
    \end{quantikz}
    .
\end{equation}
For $\exp(-i dt {B})$, one can use the circuit~\cite{Smith2019}:
\begin{equation}
    \label{eq:ZZ_unitary}
    {N}_{ij}(\gamma)
    \equiv
    e^{i \gamma \sig^{z}_{i} \sig^{z}_{j} }
    =
    \begin{quantikz}
    \lstick{$i$} & \targ{}   & \gate{R_Z(-2\gamma)}  & \targ{}   & \\
    \lstick{$j$} & \ctrl{-1} & \ghost{R_Z(-2\gamma)} & \ctrl{-1} &
    \end{quantikz}
    .
\end{equation}
For the one-dimensional TFIM, a circuit that achieves a single Trotterized time step forward is then given by (ignoring boundary conditions):
\begin{align}
    &e^{-i dt \HH(t_s)}
    \sim
    \prod_{i} {R}^x_i\left( -2 d t h(t_s) \right)
    \nonumber\\
    &\cdot
    \prod_{i\in\text{even}} {N}_{i,i+1} \left( d t J(t_s) \right)
    \prod_{i\in\text{odd}} {N}_{i,i+1} \left( d t J(t_s) \right)
    \nonumber\\
    &=
    \begin{quantikz}
    & \gate[wires=2]{{N}_{1,2}} & & \gate{{R}^x_1} & \\
    & & \gate[wires=2]{{N}_{2,3}} & \gate{{R}^x_2} & \\
    & \gate[wires=2]{{N}_{3,4}} & & \gate{{R}^x_3} & \\
    & & \gate[wires=2]{{N}_{4,5}} & \gate{{R}^x_4} & \\
    & \gate[wires=2]{{N}_{5,6}} & & \gate{{R}^x_5} & \\
    & & & \gate{{R}^x_6} & \\
    \end{quantikz}
    .
\end{align}
To avoid unnecessary idling during the application of Ising unitaries, we apply all ``odd'' (``even'') sublayers in unison.

Note that taking $A$ to be given by the transverse-field terms in the Hamiltonian and $B$ by the Ising terms is arbitrary and we could swap this choice.
However, for $d t > 0$, the Suzuki-Trotter decomposition is not invariant under the exchange of $A$ and $B$.
Thus, we have a choice to make in the operator ordering.
Here, our decision is based on the initial state used in the quench protocol, $| + \rangle^{\otimes N}$.
Acting on this state with a layer of $R_X$ rotation gates would, in principle, only contribute an unobservable global phase.
Because of this, we choose to apply Ising gates before transverse-field gates in the Trotterized circuit implementation of the time evolution operator.

\subsection{Choice of Trotter step duration}
\label{SI:trotter_step_choice}

\begin{figure*}[htbp]
    \centering
    \includegraphics[width = 0.9\textwidth]{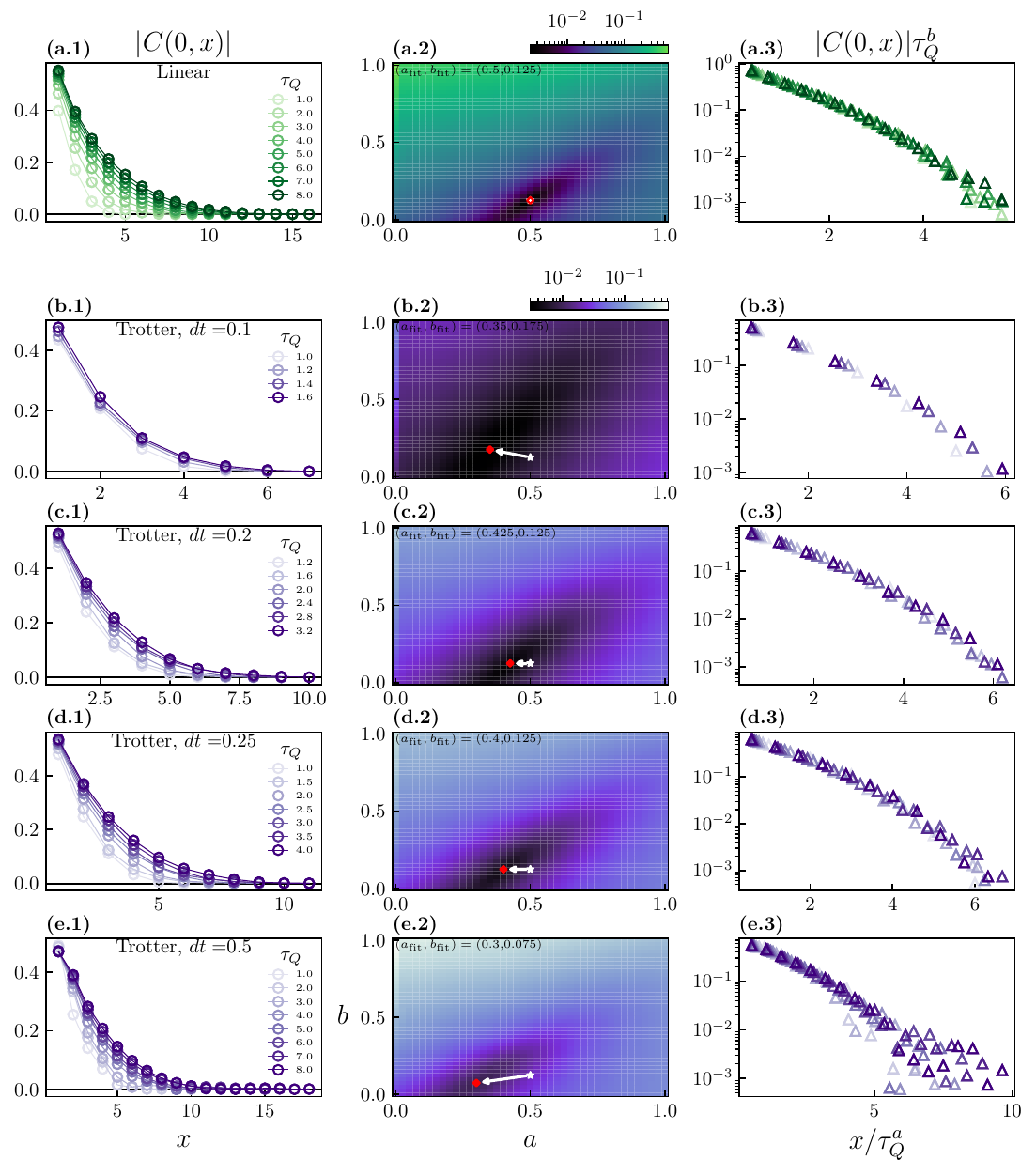}
    \caption{
    \textbf{Trotterization affects on critical scaling behavior.}
    Theoretical noiseless simulations of the equal-time connected spin-spin correlation function using $N=120$ qubits quenched to the QCP ($t=0$).
    The column layout is the same as Fig.~\ref{fig:QND_QKZ} in the main text.
    (a) Results for the continuous quench protocol in Fig.~\ref{fig:Experimental_setup}(a) with $\tau_Q = 1, 2, \dots, 8$.
    For this large finite chain, one finds excellent agreement with the thermodynamic-limit scaling exponents predicted by the QKZ mechanism.
    (b)-(e) Trotterized quenches to the QCP using Trotter-step durations $dt=0.1,0.2,0.25,0.5$ respectively with $\tau_Q = d t \cdot \mathrm{steps}$ for a varying number of steps.
    [Here, we take the number of Trotter steps to be an even number up to a maximum of 16 and require $\tau_Q \ge 1$.]
    (b) For $dt=0.1$, Trotterization errors are smaller than those in (c)-(e), but fewer quenches satisfy the imposed constraints, $1 \le \tau_Q \le 16 dt$, which leads to enhanced deviations between the QKZ scaling prediction (white star) and the fitted exponents (red diamond).
    (c) and (d) For intermediate Trotter durations, $dt=0.2, 0.25$, one finds the best agreement between QKZ scaling predictions and the fitted exponents.
    (e) For $dt=0.5$, Trotterization errors are the largest of the given data sets, and this results in the largest deviation between QKZ scaling predictions and the fitted exponents while also yielding a poor data collapse.
    }
    \label{fig:supplement_Trotter}
\end{figure*}

In Fig.~\ref{fig:supplement_Trotter} we show how the choice of Trotter step duration $dt$ affects critical scaling behavior in noiseless finite-size systems.
Anticipating the use of only a subset of available qubits on the \texttt{ibm\_fez} Heron processor for our quantum simulations, here we consider $N=120$ qubits quenched to the QCP of the TFIM in a zero-noise environment using both continuous linear quench protocols (see Fig.~\ref{fig:Experimental_setup}(a) in the main text) and Trotterized quench protocols.
For each quench, using the method outlined in Section~\ref{SI:theory_numerics}, we numerically determine the equal-time connected spin-spin correlation function, Eq.~(\ref{eq:site_averaged_connected_2pt_fn}).
In Fig.~\ref{fig:supplement_Trotter}(a), we show correlation function data using the continuous linear quench protocol with quench times $\tau_Q = 1, 2, \dots, 8$.
In this finite, but large, chain the scaling exponents that lead to the best data collapse in $C(0,x) \tau_Q^{b}$ versus $x / \tau_Q^{a}$ are in fantastic agreement with those predicted by the QKZ mechanism.
For the Trotterized quench protocols, we show results for $dt = 0.1, 0.2, 0.25, 0.5$ in Fig.~\ref{fig:supplement_Trotter}(b-e) respectively.
In principle, one would like to take $dt \ll 1$ in order to minimize errors associated with the first-order Suzuki-Trotter decomposition.
However, in studying QKZ physics, we also want to consider quenches with large quench times, $\tau_Q = dt \cdot \mathrm{steps}$, with $\mathrm{steps}$ being the number of Trotter steps to the QCP.
For smaller $dt$, a larger number of Trotter steps is required to reach the same quench time, but we must contend with the experimental reality that in present-day digital quantum computers, one can only apply a relatively small number of gate layers before hardware noise destroys quantum coherence.
Setting the maximum number of Trotter steps as $16$ in our zero-noise simulations, we find that both $dt = 0.2$ and $dt = 0.25$, Fig.~\ref{fig:supplement_Trotter}(c) and Fig.~\ref{fig:supplement_Trotter}(d) respectively, work well as a compromise between lowering Trotter error and extending $\tau_Q$.
For $dt=0.1$, Fig.~\ref{fig:supplement_Trotter}(b), fewer quenches satisfy $1 \le \tau_Q \le 16 dt$ which leads to poorer agreement between the fitted scaling exponents and the QKZ predictions. 
For $dt=0.5$, Fig.~\ref{fig:supplement_Trotter}(e), more quenches satisfy the constraints imposed on $\tau_Q$, but increased Trotter errors now lead to poor agreement between the fitted scaling exponents and the QKZ predictions along with a poor data collapse.

\subsection{Choice of sample shot and standard errors}
\label{SI:sampling}

All IBM Quantum hardware simulations reported in this work were performed with the Qiskit Runtime Sampler primitive using $2^{20}$ shots.

The number of shots was chosen so that the standard error in the measured equal-time correlation functions would be on the order of $10^{-3}$.
The standard error in each reported correlation function is proportional to $1/\sqrt{\mathrm{shots}} = 1 / 1024 \approx 9.8 \times 10^{-4}$.
For example, the standard error in the local magnetization is given by $\sqrt{(1-[\langle \sig^\alpha_i \rangle_t]^2)/\mathrm{shots}}$ (data for $\alpha=x,z$ are reported in the main text).
Whereas the standard error in the site-averaged equal-time connected spin-spin correlation function $C(t,x)$ [Eq.~(\ref{eq:site_averaged_connected_2pt_fn}) in the main text] is given by
\begin{equation}
    \sqrt{\frac{\frac{1}{N^2}\sum_{ij} \left[
    \langle \sig^z_{i} \sig^z_{i+x} \sig^z_{j} \sig^z_{j+x}\rangle_t
    -
    \langle \sig^z_{i} \sig^z_{i+x} \rangle_t
    \langle \sig^z_{j} \sig^z_{j+x} \rangle_t
    \right]}{\mathrm{shots}}}
    .
\end{equation}
In practice, the standard error for each of the measured quantities was so small compared to its expectation value that we opted not to include error bars on plots.

\subsection{Choice of error mitigation}
\label{SI:error_mitigation}

All IBM Quantum hardware simulations reported in this work were performed with the Qiskit Runtime Sampler primitive using level 3 transpiler optimization and minimal error mitigation.
As this work seeks to study inherent noise in present-day IBM Quantum hardware, only error mitigation schemes with minimal cost were utilized.
In this case, Pauli twirling and dynamical decoupling:
the Sampler options \texttt{twirling.enable\_gates}, \texttt{twirling.enable\_measure}, and \texttt{dynamical\_decoupling.enable} were all set to \texttt{True}.

\subsection{Expanded experimental data}

\begin{figure*}[htbp]
    \centering
    \includegraphics[width = 0.9\textwidth]{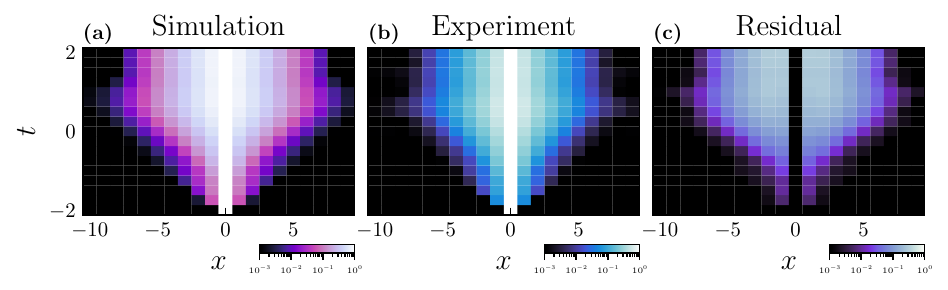}
    \caption{
    \textbf{Extended data for connected correlations during a quench of the transverse-field Ising model.}
    Dynamics of the equal-time connected spin-spin correlation function $|C(t,x)|$ for $N=20$ qubits with $16$ Trotter steps of duration $dt = 0.25$ ($-2 \le t \le 2$).
    (a) Theoretical zero-noise simulations of the Trotterized quench. 
    (b) Quantum simulations performed on \texttt{ibm\_fez}.
    The data is the same as that shown in Fig.~\ref{fig:N=20_Dynamics} of the main text.
    However, unlike Fig.~\ref{fig:N=20_Dynamics}, here we employ a logarithmic scale color map to visually amplify spacetime regions where the signal rises above the standard error.
    The lower bounds of each color map, $10^{-3}$, was chosen due to the standard error associated with the measured correlation function data.
    (c) Absolute value of the difference between experimental and theoretical data sets.
    This quantifies deviations between the quantum hardware and noiseless simulations and shows the growth of noise-induced correlations with increasing gate layers.
    }
    \label{fig:supplement_N=20_dynamics}
\end{figure*}

In this subsection, we show additional data and visualizations from our digital quantum simulations. First, in Fig.~\ref{fig:supplement_N=20_dynamics} we show correlation function dynamics using $N=20$ qubits over the interval $-\tau_Q \le t \le \tau_Q$ with 16 Trotter steps of duration $dt = 0.25$ ($\tau_Q = 2$).
This is the same data as shown in Fig.~\ref{fig:N=20_Dynamics} of the main text, but here we display the correlation function using a logarithmic scale color map to visually emphasize spacetime regions where the measured data rises above the standard error.
We again include noiseless theory simulations (left column) to contrast the hardware results (middle column), but also quantify their difference by showing the residuals between the data sets (right column).

\begin{figure*}[htbp]
    \centering
    \includegraphics[width = 0.9\textwidth]{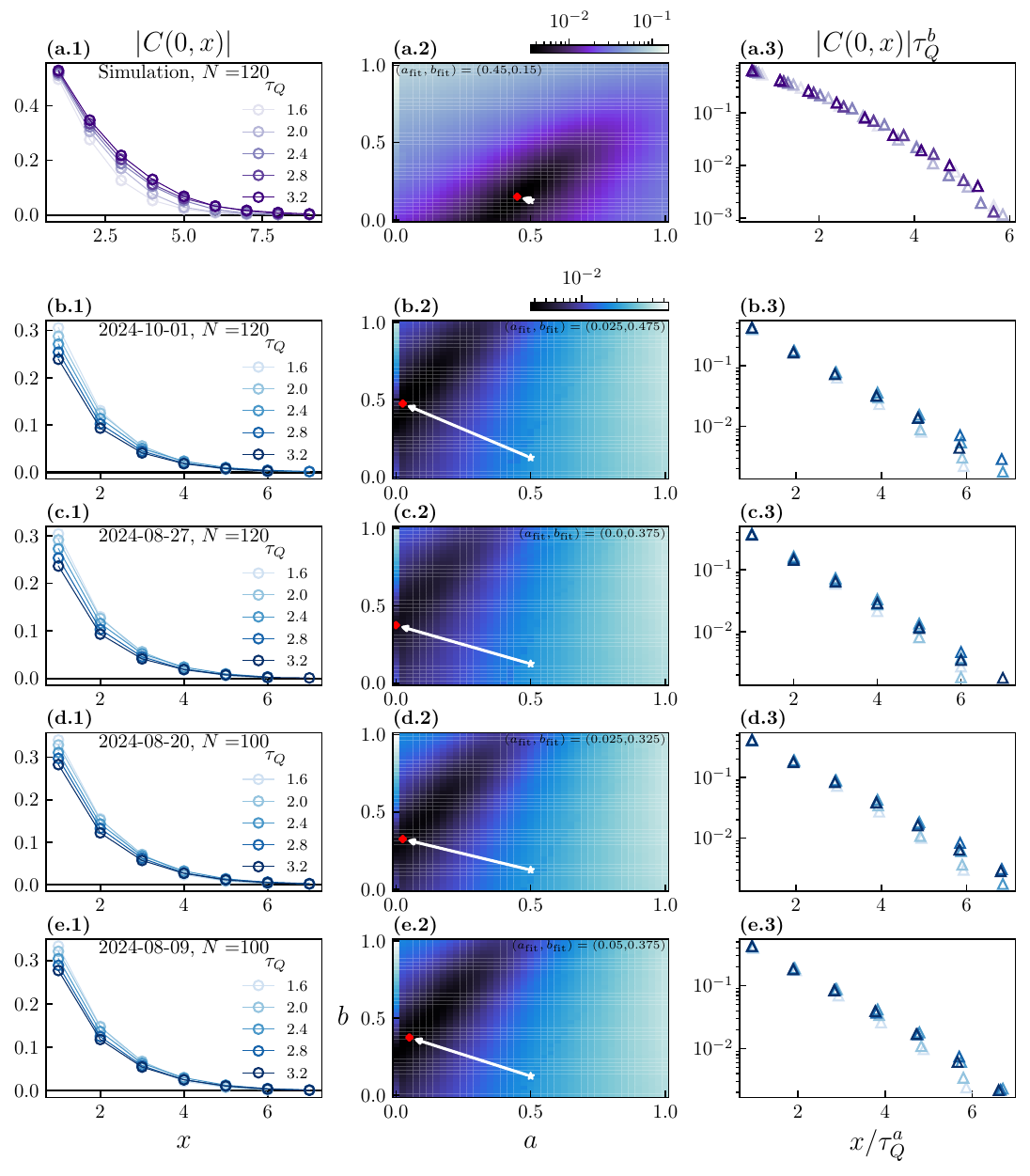}
    \caption{
    \textbf{Extended data for scaling behavior at the quantum critical point in digital quantum computers.}
    Additional experimental data from measuring the equal-time correlation function at the QCP using Trotter steps of duration $dt=0.2$.
    This image serves to supplement Fig.~\ref{fig:Anti-KZ_QCP} in the main text.
    Once again, column 1 shows the correlation function data, column 2 shows the search for data collapse scaling exponents, and column 3 shows the re-scaled data.
    (a) Zero-noise simulations of the Trotterized quench, as shown in Fig.~\ref{fig:Anti-KZ_QCP}.
    (b)-(e) Quantum simulations performed on \texttt{ibm\_fez} over various various dates from 2024-08-09 to 2024-10-01.
    Because of statistical shot noise, correlation function data below $10^{-3}$ is omitted, as is the $x=0$ point where QKZ scaling does not apply.
    [Note: we have additionally masked $x>13$ data points due to a small number of spurious correlations arising above the shot noise error at greater spatial separations.]
    As discussed in the main text, the fitted scaling exponents on the quantum hardware differ noticeably from noiseless simulations and QKZ predictions.
    Yet the collapse exhibited in the re-scaled data is consistent with universal behavior shaped by the decoherent environment.
    }
    \label{fig:supplement_QCP}
\end{figure*}

Second, in Fig.~\ref{fig:supplement_QCP} we show further experimental data when quenching the system to the QCP using Trotter steps of duration $dt=0.2$.
This image contains quenches simulated with \texttt{ibm\_fez} over multiple dates from 2024-08-09 to 2024-10-01 using both $N=100$ and $N=120$ qubits.
The data from 2024-10-01 and 2024-08-20 is what appears in Fig.~\ref{fig:Anti-KZ_QCP} of the main text.
For each quantum simulation, the best-fit scaling exponents (red diamonds in column 2) share a consistent trend of $a_\mathrm{fit}$ decreasing and $b_\mathrm{fit}$ increasing when compared to the values predicted by the QKZ mechanism (white stars in column 2).

\end{document}